\documentclass[twocolumn,floatfix,superscriptaddress,showpacs,showkeys]{revtex4}
\usepackage{amsmath}
\usepackage{graphicx}
\usepackage{dcolumn}
\usepackage{bm}
\usepackage{booktabs}
\usepackage{amssymb}
\usepackage{multirow}
\usepackage{makecell}
\usepackage{textcomp}
\usepackage{subfigure}
\usepackage{phonetic}
\usepackage{extarrows}
\usepackage{color}
\usepackage{float}
\usepackage[colorlinks,citecolor=blue,linkcolor=blue]{hyperref}
\makeatletter
\newcommand{\rmnum}[1]{\romannumeral #1}
\newcommand{\Rmnum}[1]{\expandafter\@slowromancap\romannumeral #1@}
\makeatother

\begin{document}


\title{Robust interface-state laser in non-Hermitian micro-resonator arrays}

\author{Lu Qi}
\affiliation{School of Physics, Harbin Institute of Technology, Harbin, Heilongjiang 150001, China}
\author{Guo-Li Wang}
\affiliation{School of Physics, Harbin Institute of Technology, Harbin, Heilongjiang 150001, China}
\author{Shutian Liu}
\email{stliu@hit.edu.cn}
\affiliation{School of Physics, Harbin Institute of Technology, Harbin, Heilongjiang 150001, China}
\author{Shou Zhang}
\email{szhang@ybu.edu.cn}
\affiliation{School of Physics, Harbin Institute of Technology, Harbin, Heilongjiang 150001, China}
\affiliation{Department of Physics, College of Science, Yanbian University, Yanji, Jilin 133002, China}
\author{Hong-Fu Wang}
\email{hfwang@ybu.edu.cn}
\affiliation{Department of Physics, College of Science, Yanbian University, Yanji, Jilin 133002, China}


\date{\today}

\begin{abstract}
We propose a scheme to achieve the analogous interface-state laser by dint of the interface between the two intermediate-resonator-coupled non-Hermitian resonator chains. We find that, after introducing the couplings between the two resonator chains and the intermediate resonator at the interface, the photons of the system mainly gather into the three resonators near the intermediate resonator. The phenomenon of the photon gathering towards the certain resonators is expected to construct the photon storage and even the laser generator. 
We reveal that the phenomenon is induced via the joint effect between the isolated intermediate resonator and two kinds of non-Hermitian skin effects. 
Specially, we investigate the interface-state laser in topologically trivial non-Hermitian resonator array in detail. We find that the pulsed interface-state laser can be achieved accompanying with the intermittent proliferation of the photons at the intermediate resonator when an arbitrary resonator is excited. Also, we reveal that the pulsed interface-state laser in the topologically trivial non-Hermitian resonator array is immune to the on-site defects in some cases, whose mechanism is mainly induced by the nonreciprocal couplings instead of the protection of topology. Our scheme provides a promising and excellent platform to investigate interface-state laser in the micro-resonator array.  
\end{abstract}

\maketitle


\section{\label{sec.1}Introduction}
Topological insulator~\cite{hasan2010colloquium,qi2011topological,chiu2016classification,bansil2016colloquium}, a new kind of state of the matters, has attracted a growing number of the interest among the physics community since it owns many distinctively novel properties. One of the most well-known of these properties is that the topological insulator has the conducting boundary or surface states at the interface and the insulating bulk states in the bulk simultaneously~\cite{hasan2010colloquium,qi2011topological,bansil2016colloquium}. These conducting boundary or surface states originate from the topology of the topological insulator, namely, the topological materials on both sides of the interface are topologically inequivalent~\cite{hasan2010colloquium,qi2011topological,bansil2016colloquium,wray2011topological}. More specifically, these conducting boundary or surface states appear at the interface between the topologically nontrivial matter (such as the topological insulator) and the topologically trivial matter (such as the usual insulator). It has great significance for the investigation of these conducting boundary or surface states due to its protection of the topology and numerous potential applications in quantum information processing~\cite{dlaska2017robust,aasen2016milestones,Sarma2015}. Thus, the growing efforts have been devoted to the context of these interface states, including the direct observation of the interface state between topological and normal insulators~\cite{berntsen2013direct}, the nontrivial interface states between two topological insulators~\cite{rauch2013nontrivial}, the inverse spin-galvanic effect~\cite{garate2010inverse} and the boundary states~\cite{eremeev2015interface} at the interface between the topological insulator and the ferromagnetic insulator, the multiple topologically-protected interface states in the bulk silicene~\cite{wang2014multiple}, the topological interface in ultracold atomic gases~\cite{borgh2012topological} and superconducting qubits~\cite{jiang2011interface}, etc.

Recently, the non-Hermitian extensions~\cite{malzard2015topologically,bender2007making,diehl2011topology,san2016majorana,lee2014heralded,alvarez2018non,ding2016emergence} of the topological insulator also have attracted increasing attention both in parity-time ($\mathcal{PT}$) symmetric topological system~\cite{esaki2011edge,zeuner2015observation,ghosh2012note,zhu2014pt,yuce2015topological,weimann2017topologically,hu2011absence,xing2017spontaneous} and the non-Hermitian topological system with nonreciprocal couplings~\cite{lee2016anomalous,yao2018edge,yao2018non,kunst2018biorthogonal,jin2019bulk,chen2019finite,edvardsson2019non,liu2019second,turker2019open,yuce2019topological,yokomizo2019non,lee2019hybrid}. The existence of the imaginary energy spectrum and the appearance of the non-Hermitian skin effect~\cite{yao2018edge,yao2018non} are the most prominent features in these non-Hermitian topological systems. Especially, the interface states in these non-Hermitian topological systems also have been extensively investigated, such as the topologically protected interface and the bound states in $\mathcal{PT}$-symmetric topological systems~\cite{yuce2018edge,pan2018photonic,weimann2017topologically,jones2008interface,weimann2015parity}, the interface states in nonreciprocal higher-order topological metals~\cite{ezawa2019non}, the defect states in nonreciprocal optical waveguide~\cite{bosch2019non}, the robust propagation of the electromagnetic waves at the defect interface~\cite{longhi2015non,longhi2015robust,gangaraj2018topological}, the detection of the topological boundary states and the non-Hermitian skin effect~\cite{longhi2019probing}, etc.      

In this paper, enlightened by the above investigations, we propose a scheme to implement the robust interface-state laser based on the interface states in the non-Hermitian micro-resonator array. We show that, based on a non-Hermitian micro-resonator array composed by two intermediate-resonator-coupled resonator chains with the opposite nonreciprocal coupling configurations, the photons of the resonator array mainly gather into the intermediate resonator at the interface and the two resonators around the intermediate resonator. We demonstrate that, the gathering of the photons towards the certain resonators is caused by the  joint effect between the isolated intermediate resonator and two kinds of non-Hermitian skin effects, in which one type of non-Hermitian skin effect originates from the original two non-Hermitian resonator chains while another type of non-Hermitian skin effect originates from the two new resonator chains induced by the couplings between the two original resonator chains and the intermediate resonator. The phenomenon that the photons gather towards the certain resonator has various potential applications in photon storage device and the laser generator device. We show that, via designing the nonreciprocal couplings of the resonator array appropriately, the photons will gather into the resonator at the interface intermittently with the developing of the time. It means that we can hopefully realize the pulsed interface-state laser. Especially, we investigate the pulsed interface-state laser based on the topologically trivial non-Hermitian micro-resonator array in detail. We find that the photons are intermittently gathered into the intermediate resonator at the interface corresponding to an arbitrary excitation of the resonator. Dramatically, the interface-state pulsed laser in the topologically trivial non-Hermitian resonator array is immune to the on-site defects due to the property of the photonic robust propagation induced by the nonreciprocal couplings. After scanning the certain resonator with a range of external driving frequency, the numerical simulations of the output detection spectrum reveal that the pulsed interface-state laser can be realized when an arbitrary resonator is excited. Thus, our scheme provides the new opportunities towards the non-Hermitian laser generator both in theory and in experiment. 

The paper is organized as follows: In Sec.~\ref{sec.2}, we propose a non-Hermitian resonator array composed by two resonator chains with opposite nonreciprocal coupling configurations and investigate the gathering effect of the photons. In Sec.~\ref{sec.3}, we focus on the case of the interface-state pulsed laser in a topological trivial non-Hermitian resonator array without the energy gap and analyze the robustness of the interface-state pulsed laser on the on-site defects. Finally, a conclusion is given in Sec.~\ref{sec.4}.
\begin{figure}
	\centering
	\includegraphics[width=0.95\linewidth]{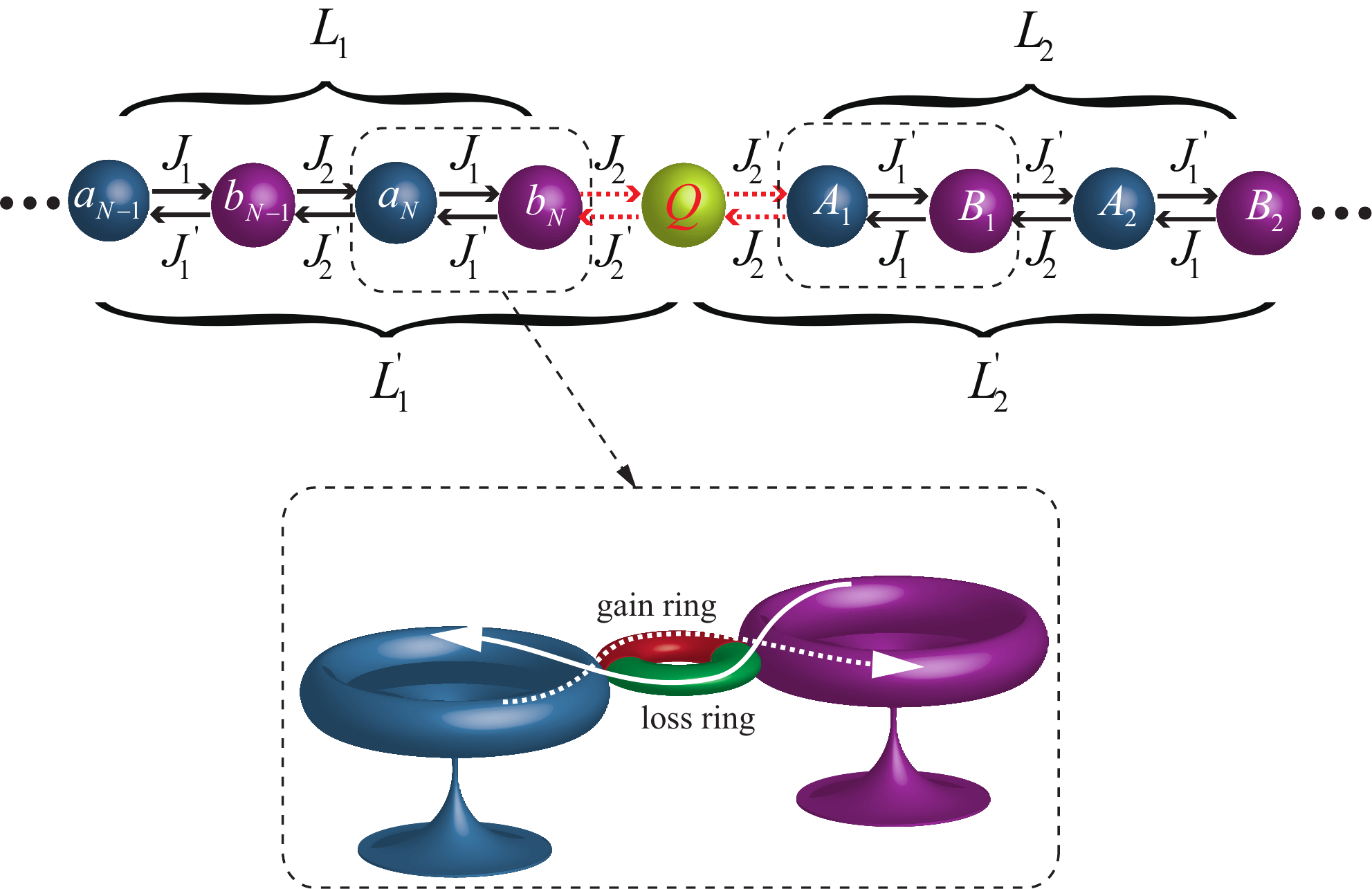}\\
	\caption{Schematic of the 1D non-Hermitian micro-resonator array. The resonator array contains two resonator chains $L_{1}$ and $L_{2}$, in which the two resonator chains both have nonreciprocal intra-cell and inter-cell couplings. The two resonator chains $L_{1}$ and $L_{2}$ couple with each other via the intermediate resonator $Q$, with the nonreciprocal coupling strengths $J_{2}$ and $J_{2}^{'}$. The nonreciprocal coupling between the two adjacent micro-resonators can be realized via an auxiliary half perimeter microring with gain in the upper half perimeter and loss in the bottom half.}\label{fig1}
\end{figure}

\section{\label{sec.2}The gathering effect of the photons in non-Hermitian topologically nontrivial resonator array} 
\begin{figure}
	\centering
	\subfigure{\includegraphics[width=0.8\linewidth]{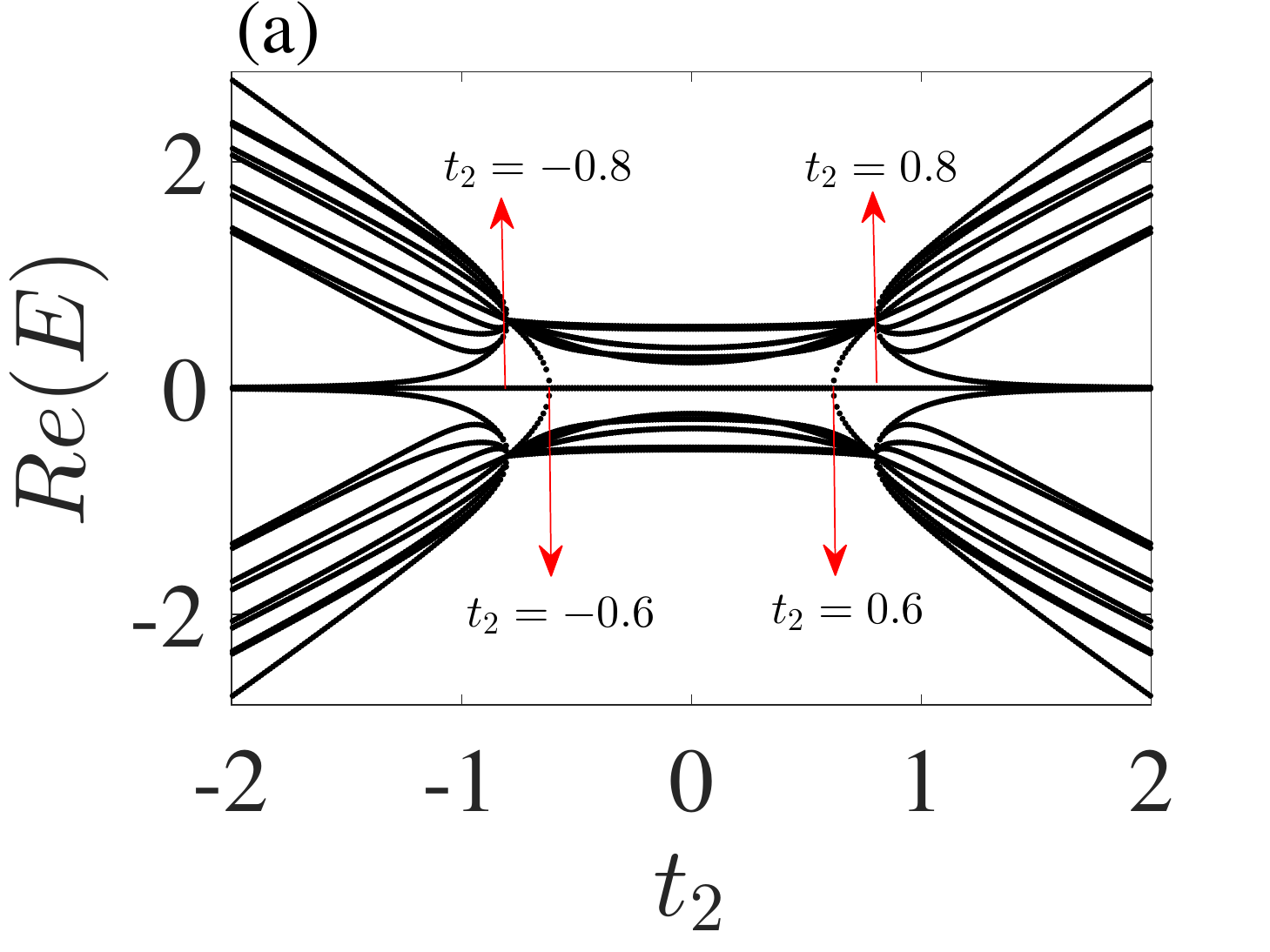}}

	\subfigure{\includegraphics[width=0.8\linewidth]{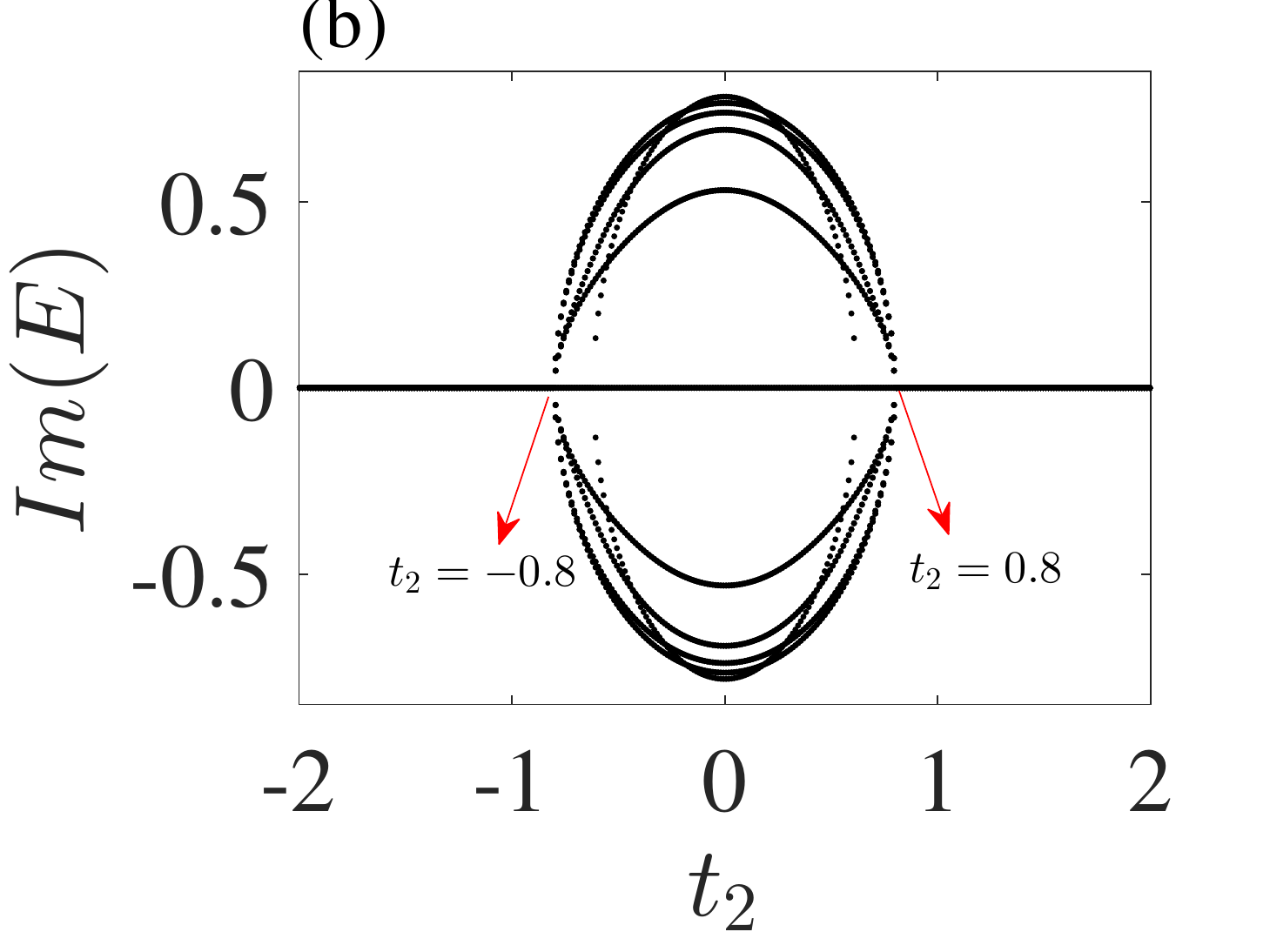}}
	
	\subfigure{\includegraphics[width=0.8\linewidth]{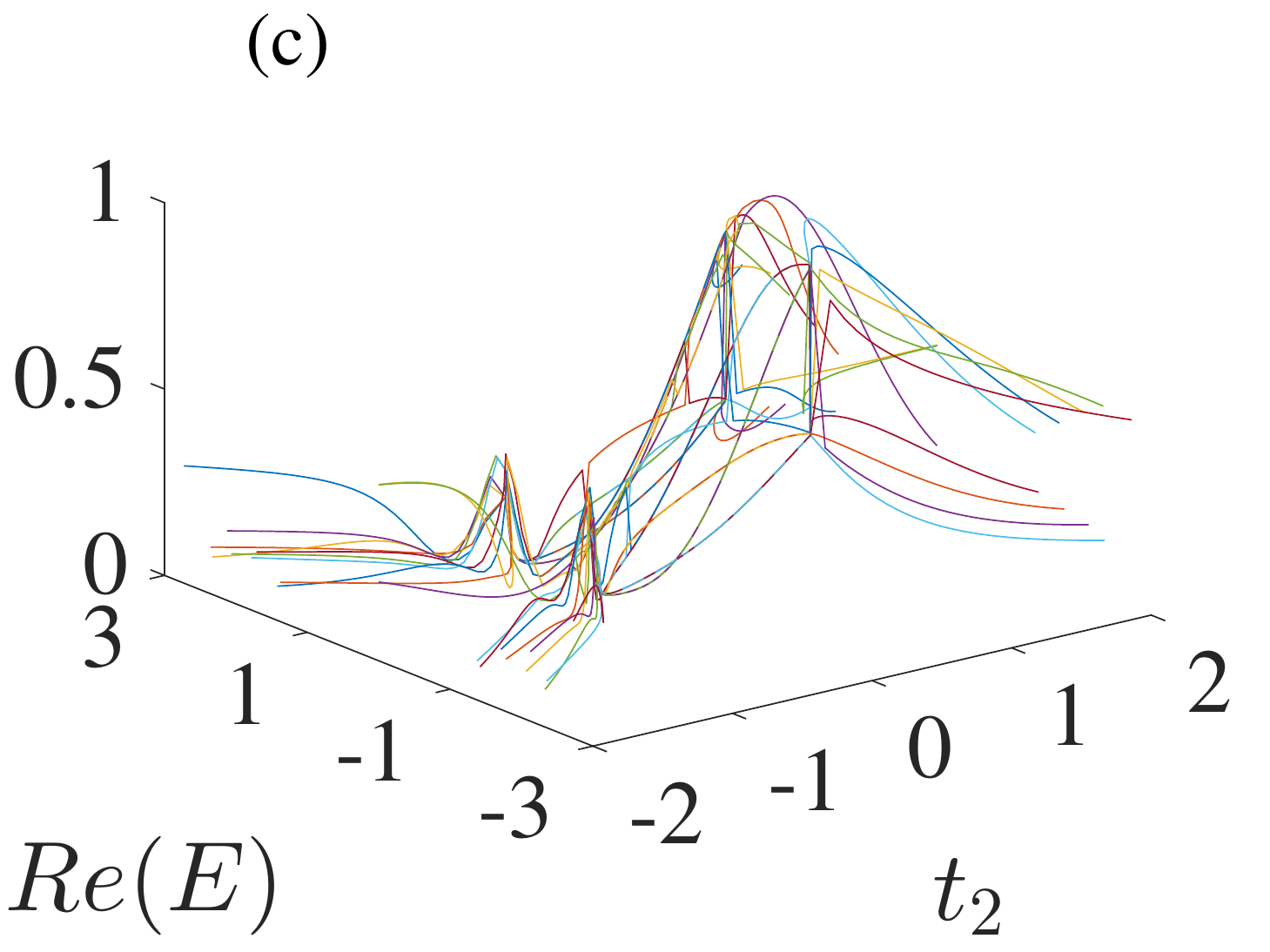}}
	
	\subfigure{\includegraphics[width=0.8\linewidth]{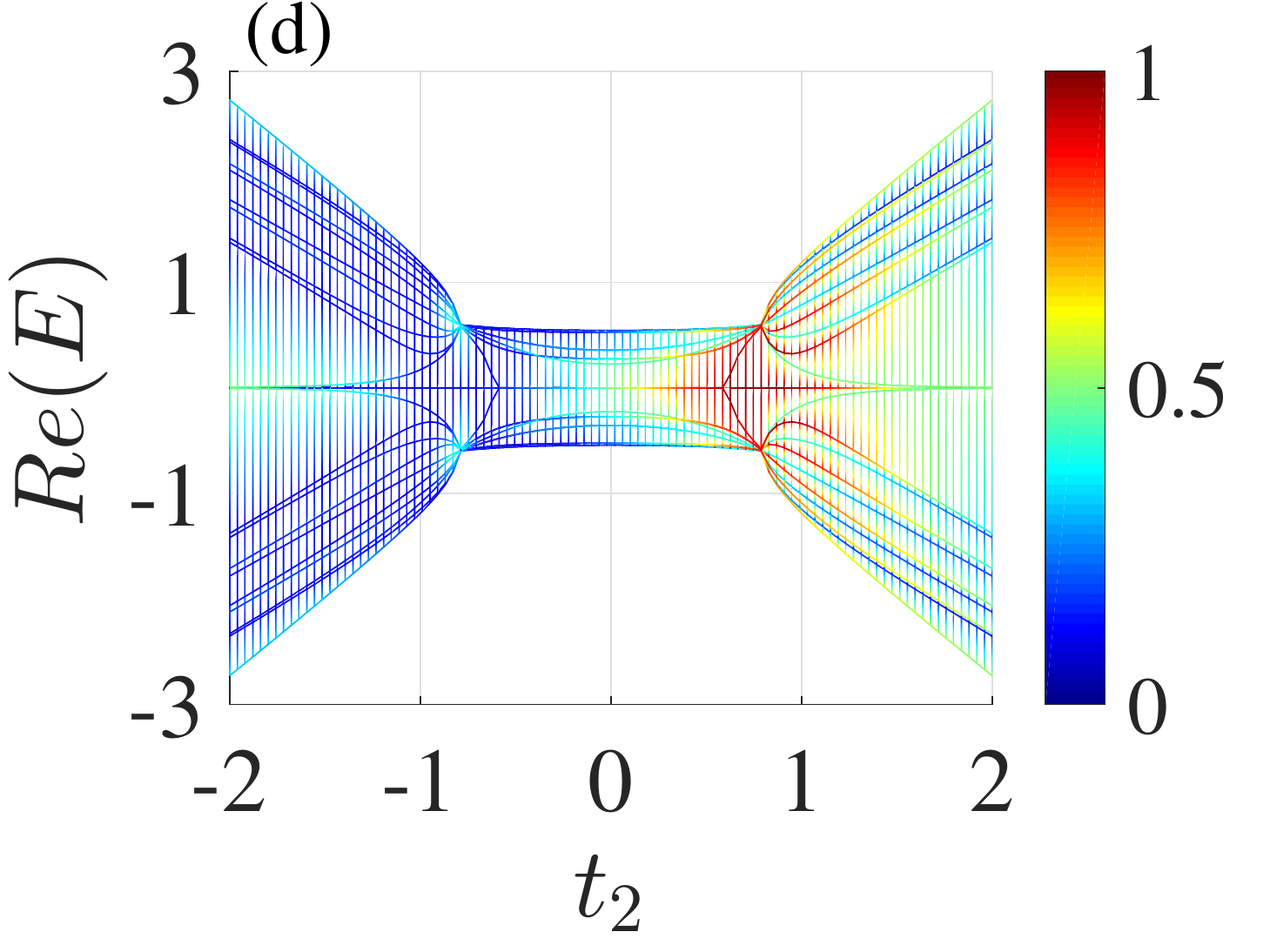}}
	\caption{The energy spectrum and localization of the system. (a) The real part of the energy spectrum versus the parameter $t_{2}$, in which the energy gap of system is divided into three parts. (b) The imaginary part of the energy spectrum versus the parameter $t_{2}$, in which the system corresponds to a pure real energy spectrum when $t_{2}>\delta$ and $t_{2}<-\delta$. (c) The inverse participation ratio (IPR) versus $t_{2}$ for each real part of energy level with $\mathrm{IPR}=\sum_{j=1}^{N}|\Psi_{n,j}\rangle^{4}/(\langle \Psi_{n}|\Psi_{n}\rangle)^{2}$, in which $j$ represents the lattice index and $|\Psi_{n}\rangle$ is the $n$th eigenstate of the system. The finite IPR corresponds to a localized state. (d) The vertical view of IPR in (c). Other parameters take $t_{1}=1$, $\delta=0.8$, and $L_{1}=L_{2}=10$. We set $t_{1}$ as the unit of energy.}\label{fig2}
\end{figure} 
\begin{figure*}
	\centering
	\subfigure{\includegraphics[width=0.49\linewidth]{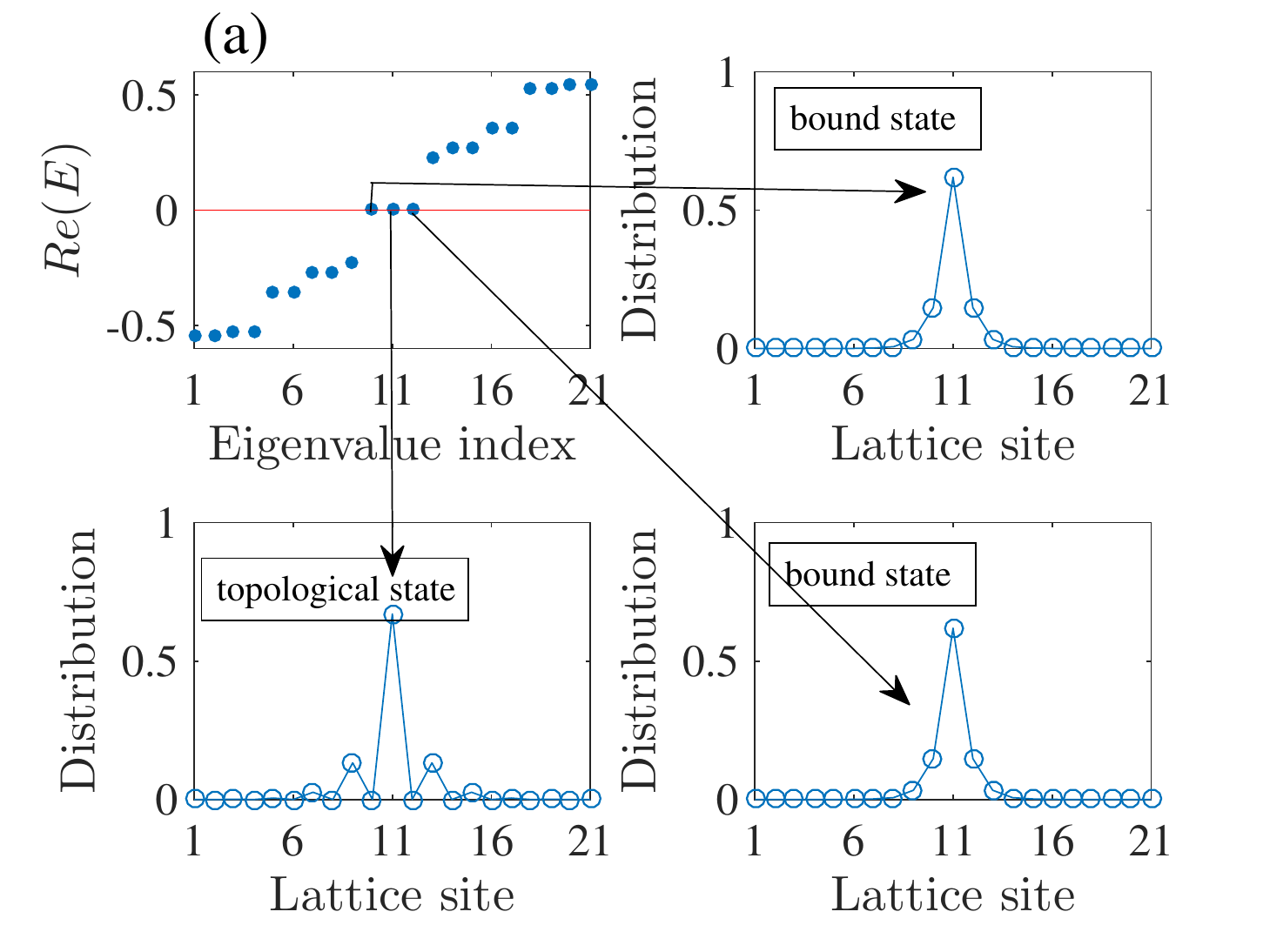}}
	\subfigure{\includegraphics[width=0.49\linewidth]{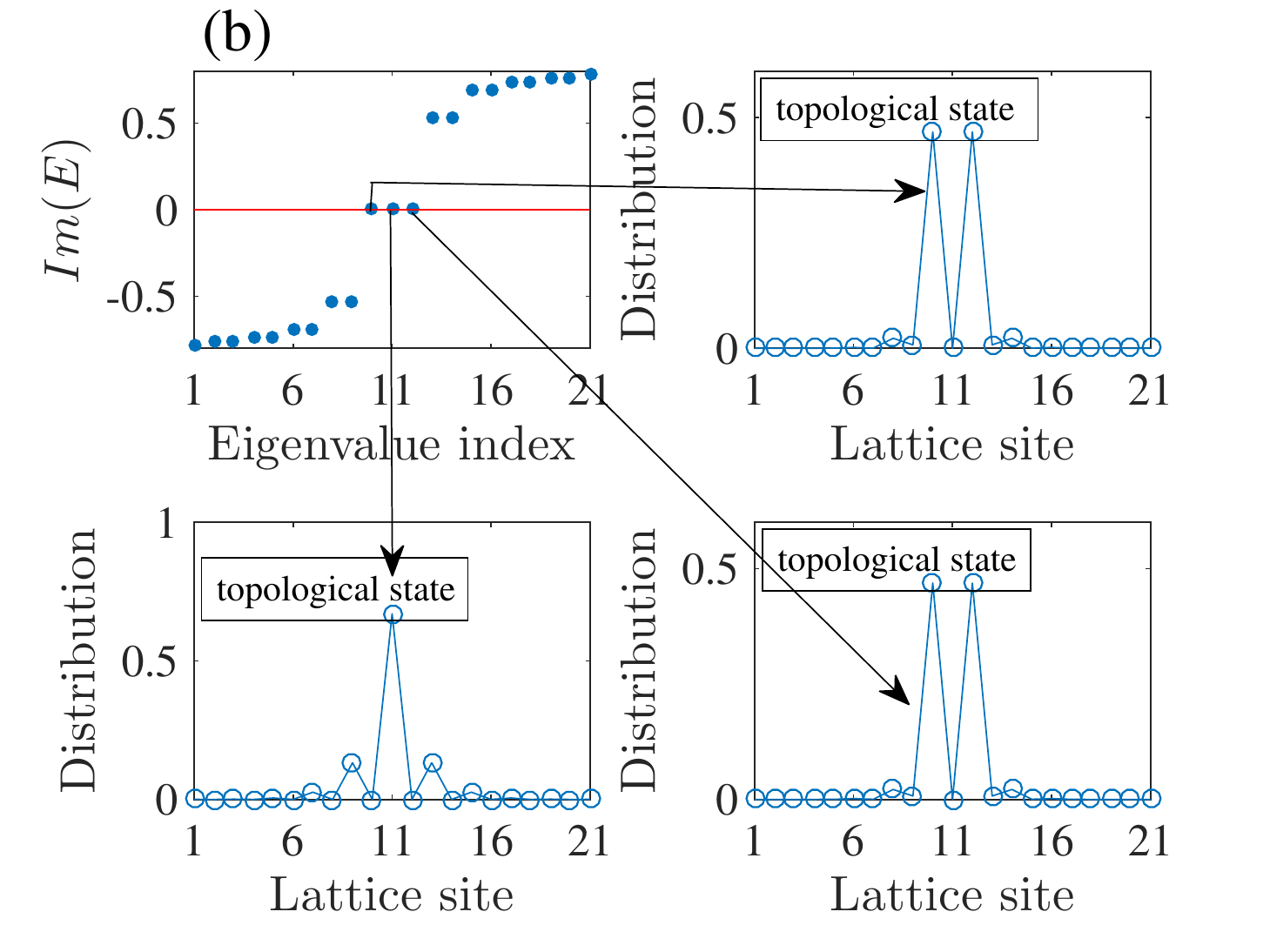}}
	
	\subfigure{\includegraphics[width=0.49\linewidth]{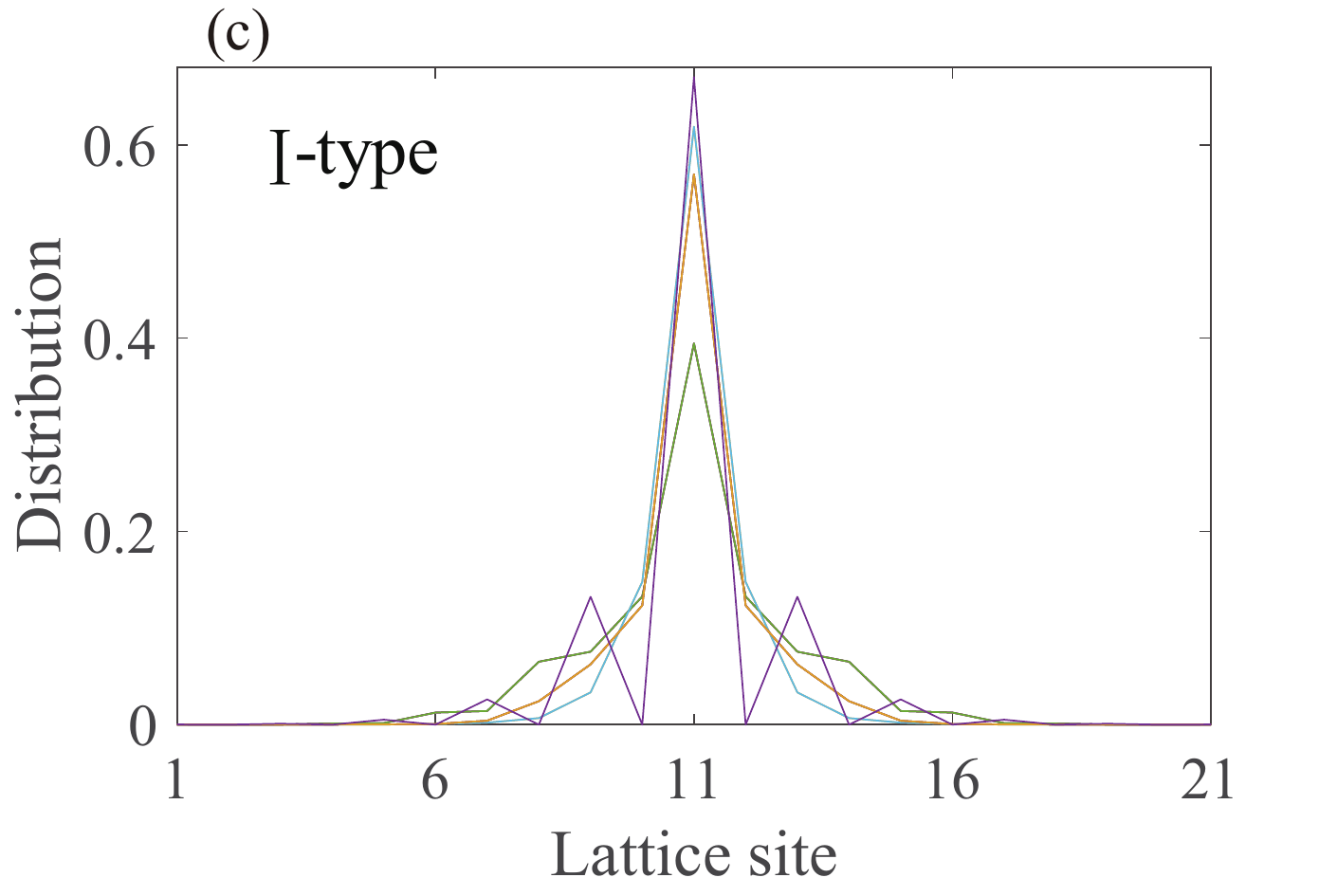}}
	\subfigure{\includegraphics[width=0.49\linewidth]{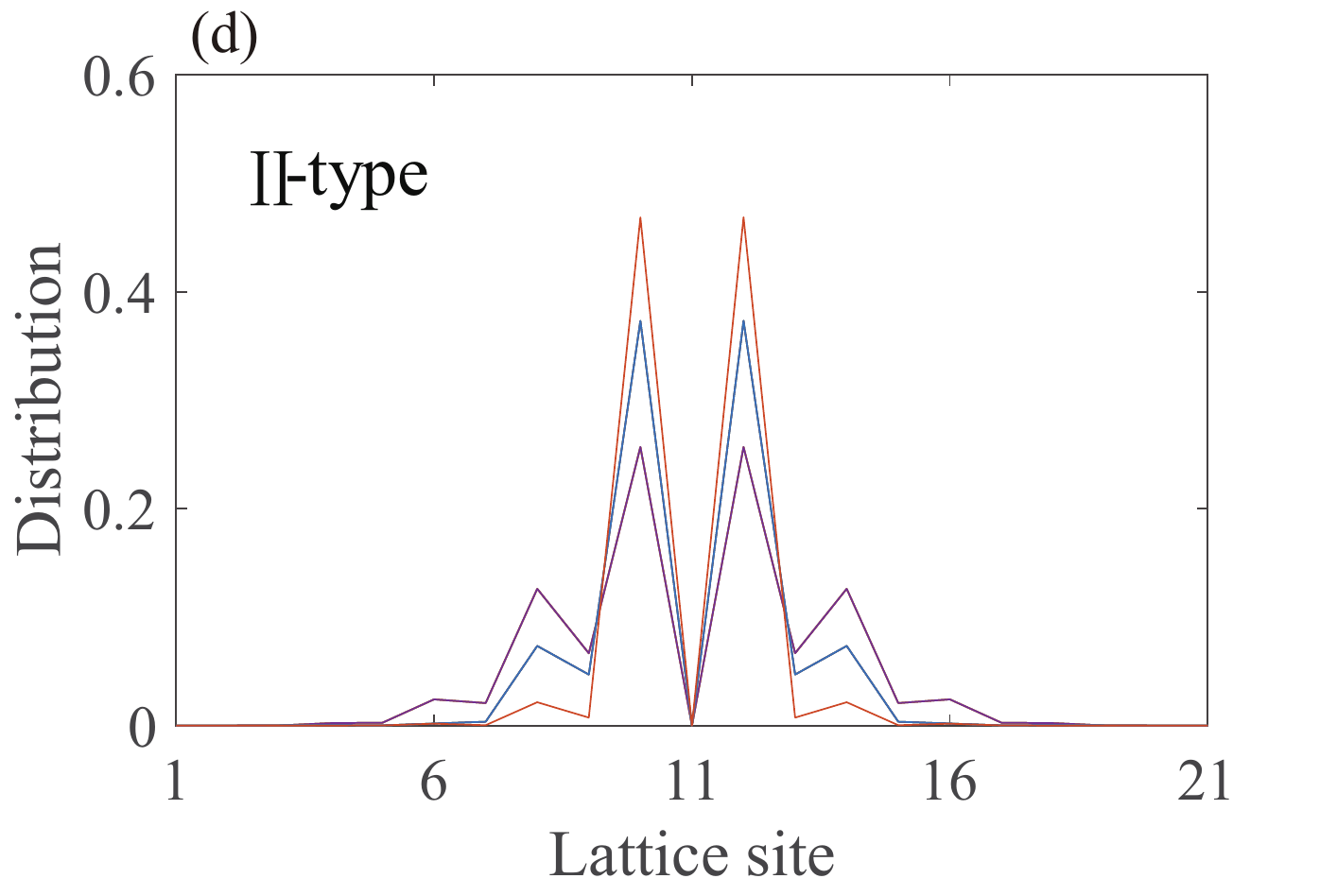}}	
	\caption{The energy spectra and the distributions of the zero energy modes. (a) The real part of the energy spectrum and the distributions of the three degenerate zero energy modes. (b) The imaginary part of the energy spectrum and the distributions of the three degenerate zero energy modes. Other parameters take $t_{1}=1$, $\delta=0.8$, $t_{2}=0$, and $L_{1}=L_{2}=10$. We set $t_{1}$ as the unit of energy.}\label{fig3}
\end{figure*} 
\begin{figure}
	\centering
	\subfigure{\includegraphics[width=0.48\linewidth]{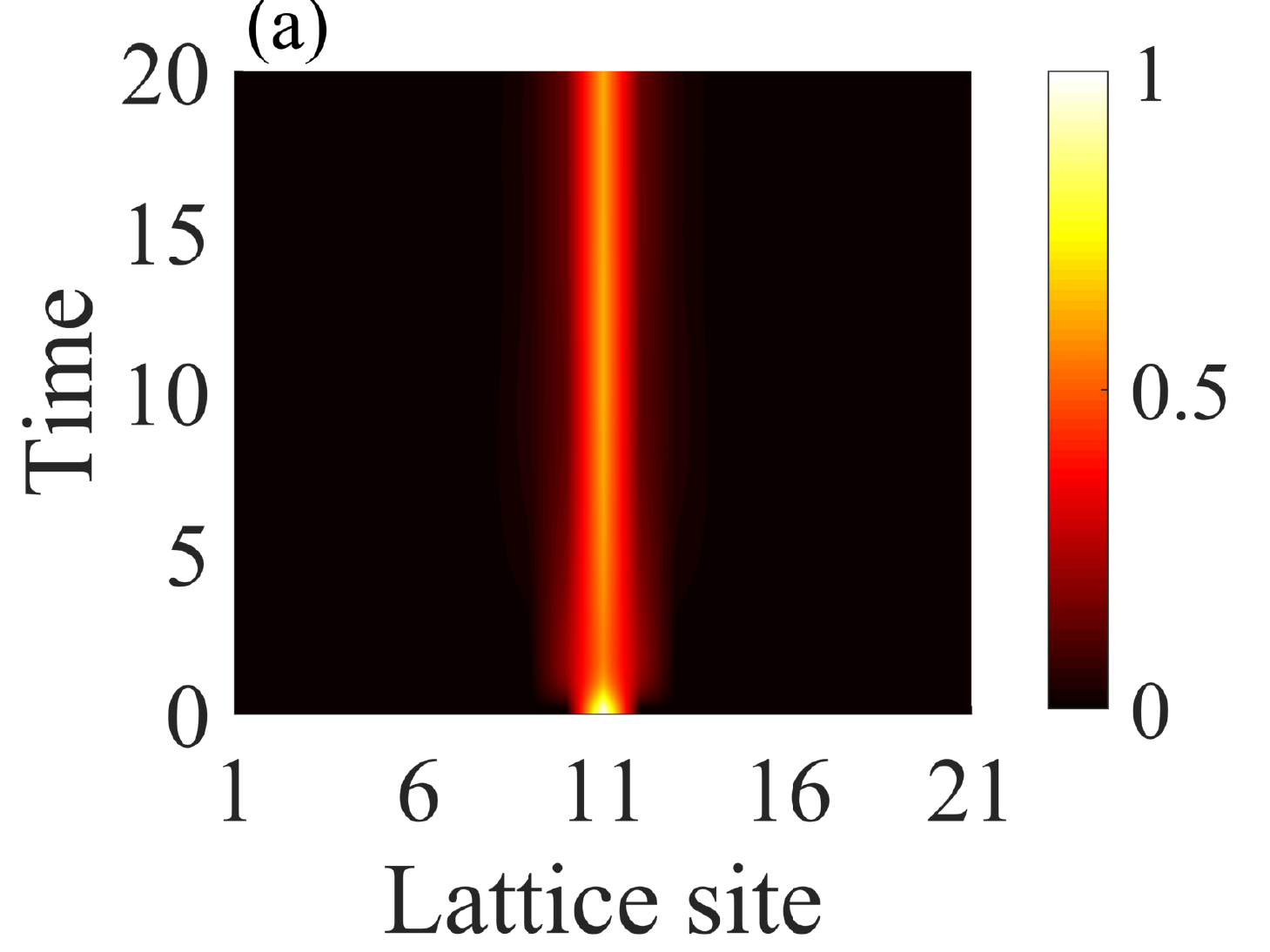}}
	\subfigure{\includegraphics[width=0.48\linewidth]{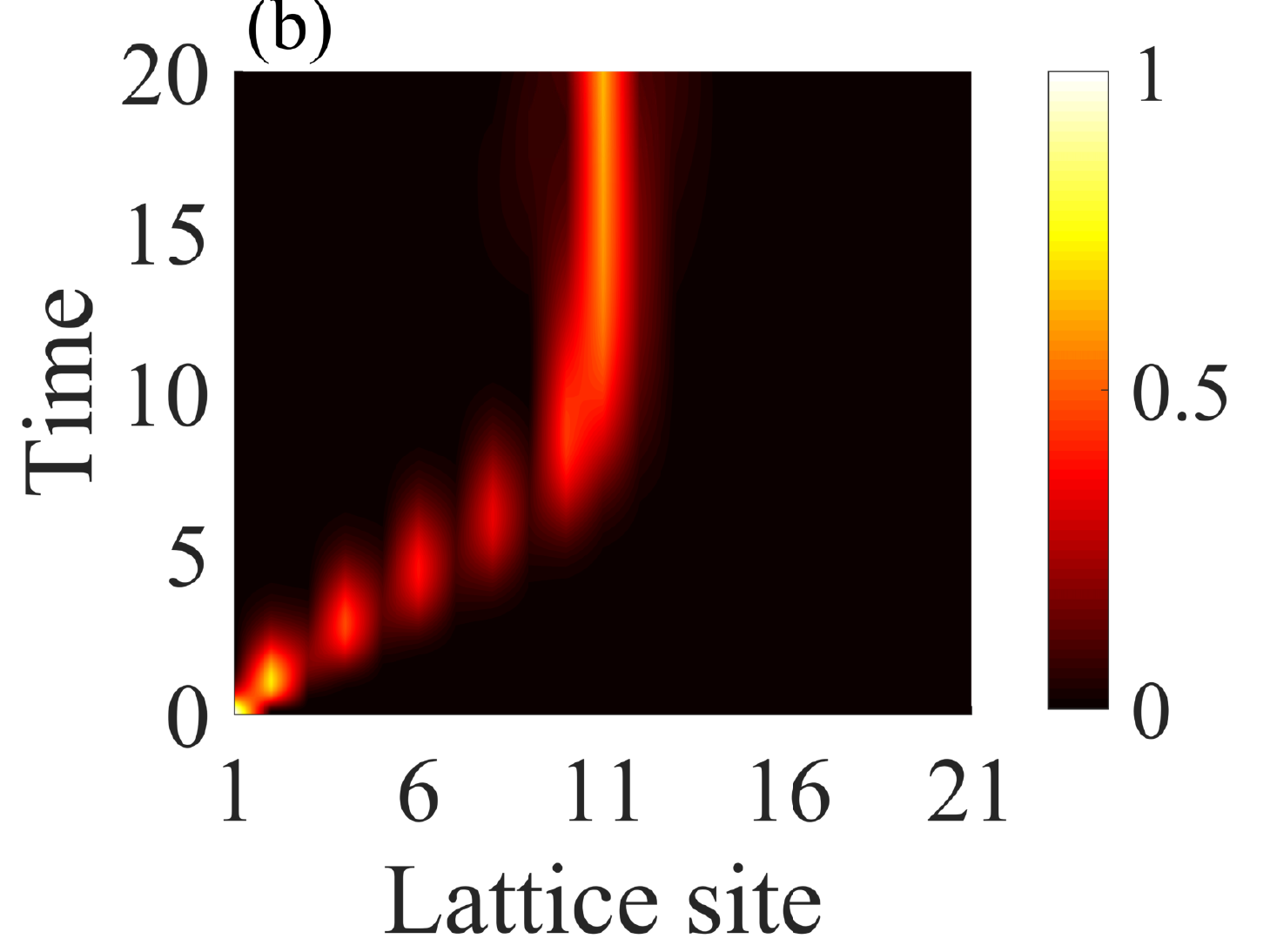}}
	
	\subfigure{\includegraphics[width=0.48\linewidth]{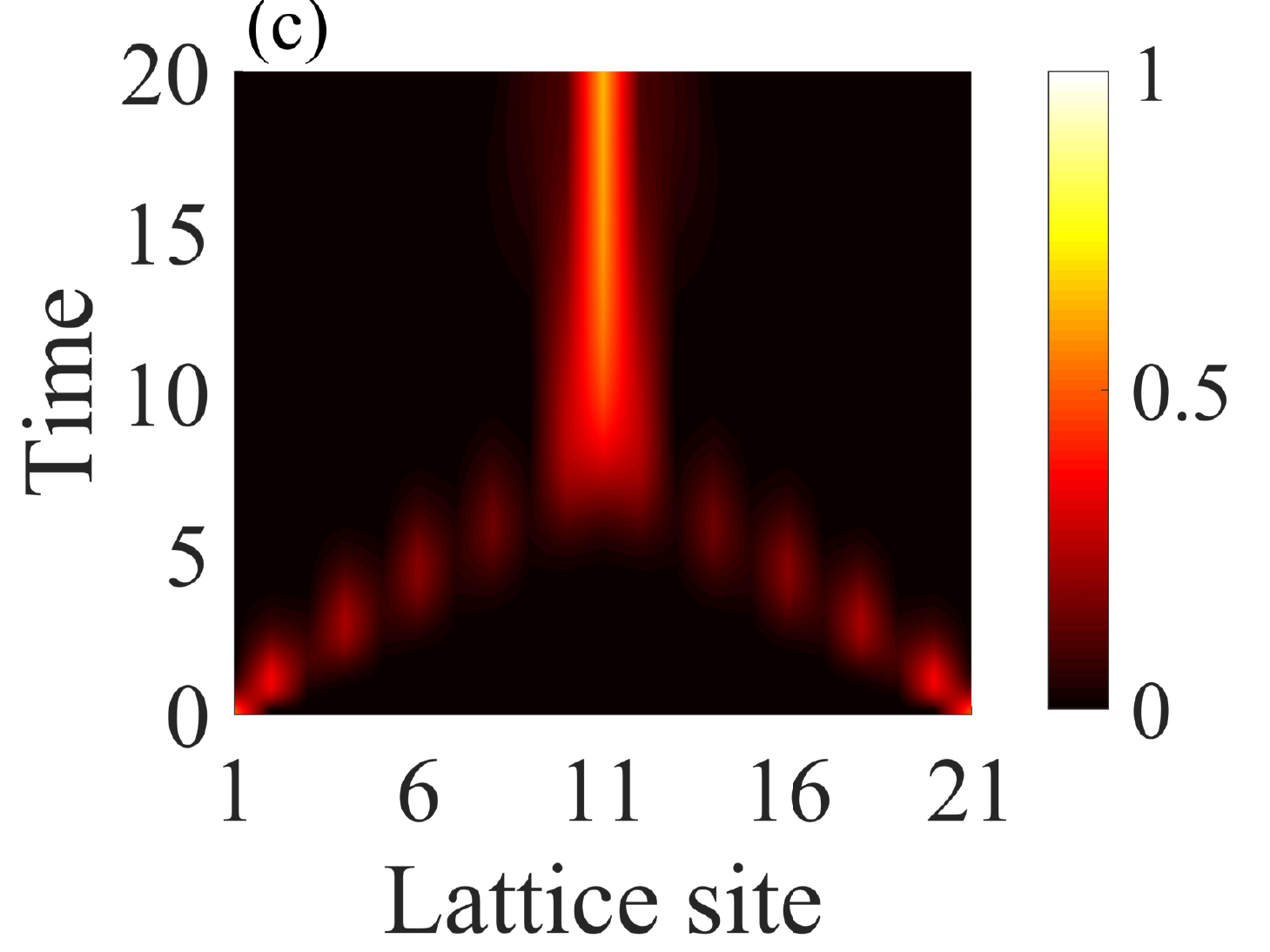}}
	\subfigure{\includegraphics[width=0.48\linewidth]{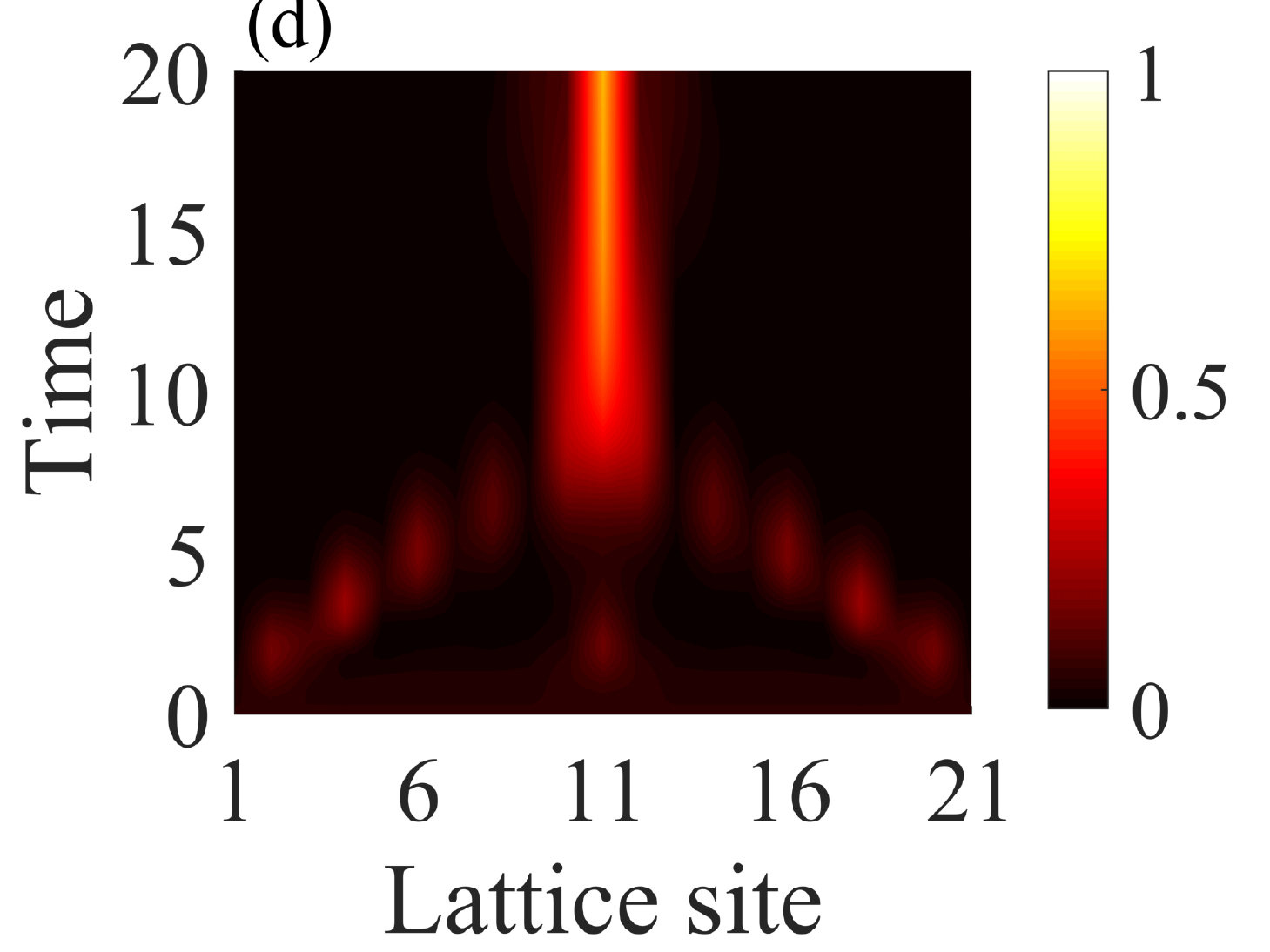}}
	\caption{The interface-state laser in the non-Hermitian resonator array when $L_{1}=L_{2}=10$. (a) The evolution of the photons when the auxiliary resonator $Q$ is excited. (b) The evolution of the photons when the first resonator is excited. (c) The evolution of the photons when the first and the last resonators are excited. (d) The evolution of the photons when all of the resonators are excited. Other parameters take $t_{1}=1$, $\delta=0.8$ and $t_{2}=0$. We set $t_{1}$ as the unit of energy.}\label{fig4}
\end{figure}    
\begin{figure*}
	\centering
	\subfigure{\includegraphics[width=0.49\linewidth]{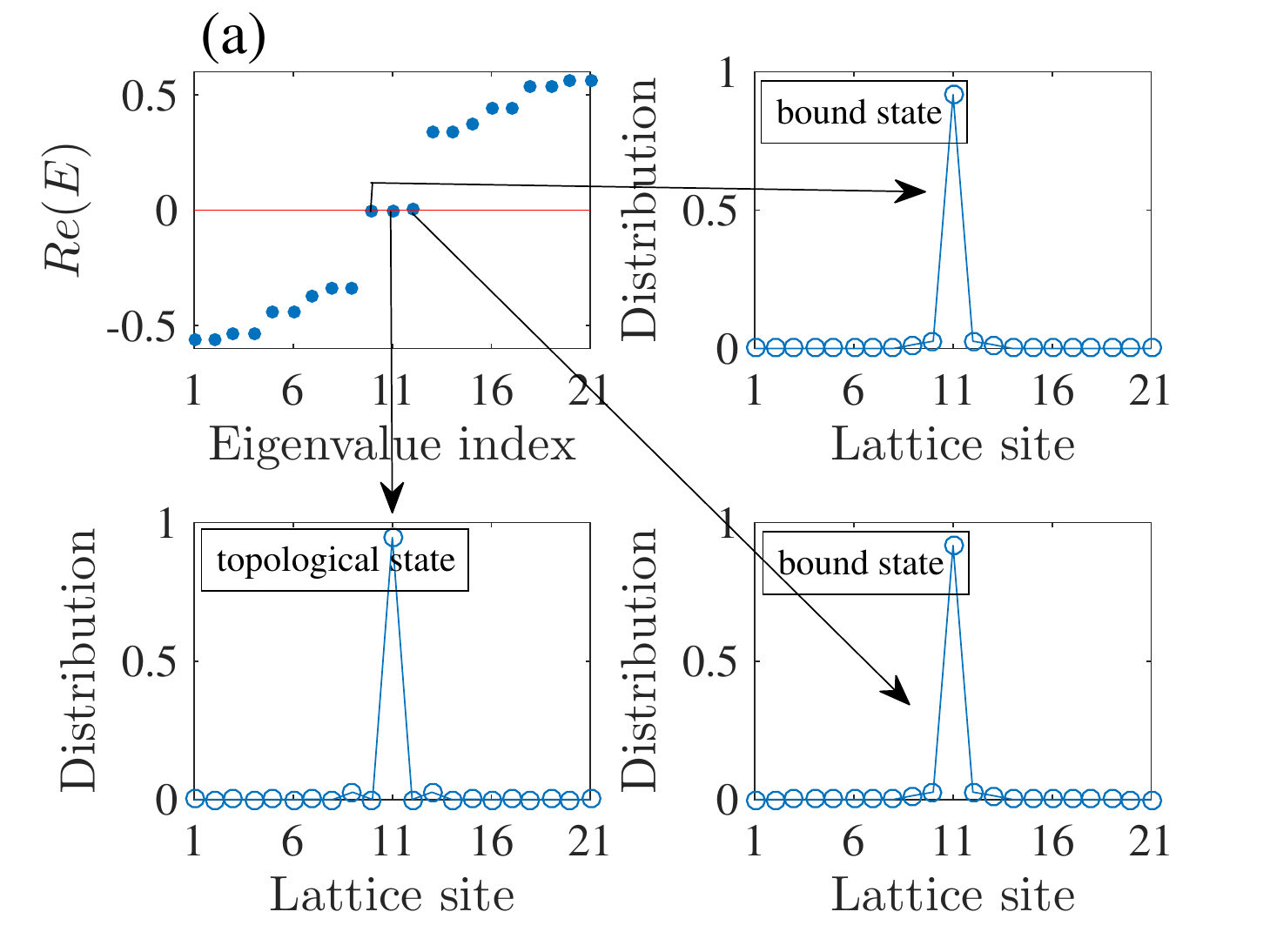}}
	\subfigure{\includegraphics[width=0.49\linewidth]{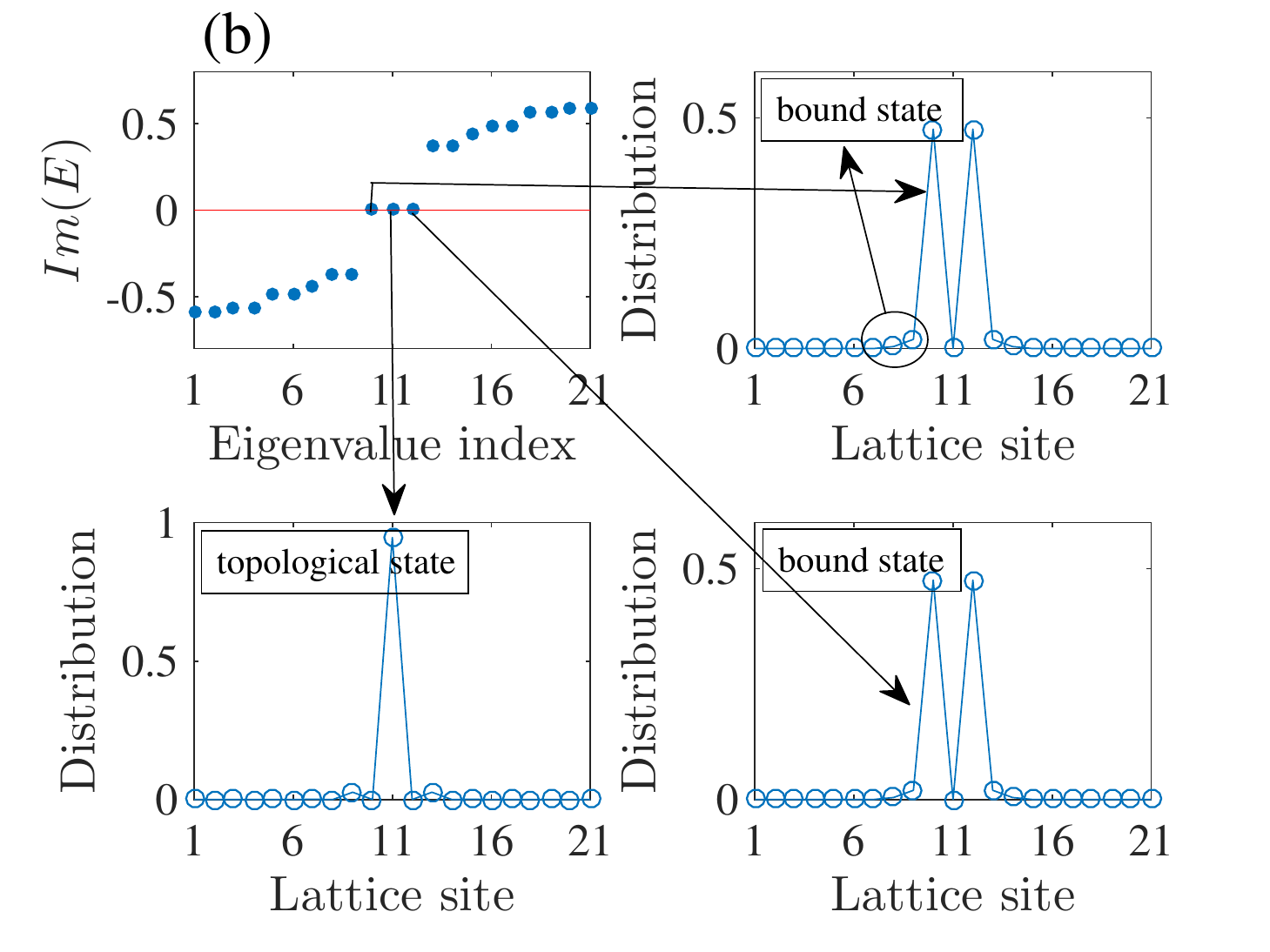}}	
	\caption{The energy spectra and the distributions of the zero energy modes. (a) The real part of the energy spectrum and the distributions of the three degenerate zero energy modes. (b) The imaginary part of the energy spectrum and the distributions of the three degenerate zero energy modes. Other parameters take $t_{1}=1$, $\delta=0.8$, $t_{2}=0.5$, and $L_{1}=L_{2}=10$. We set $t_{1}$ as the unit of energy.}\label{fig5}
\end{figure*}

\subsection{\label{sec.A}Model and theoretical analysis}

We consider a 1D non-Hermitian micro-resonator array consisting of two resonator chains $L_{1}$ and $L_{2}$, in which the two chains couple with each other via the auxiliary resonator $Q$, as shown in Fig.~\ref{fig1}. In this array, the two resonator chains $L_{1}$ and $L_{2}$ both have the nonreciprocal intra-cell and inter-cell couplings. Then, the system can be described by the Hamiltonian $H=H_{L_{1}}+H_{L_{2}}+H_{\mathrm{Link}}$, with
\begin{widetext}
\begin{eqnarray}\label{e01}
H_{L_{1}}&=&\sum_{n}\left[J_{1}b_{n}^{\dag}a_{n}+J_{1}^{'}a_{n}^{\dag}b_{n}+J_{2}a_{n+1}^{\dag}b_{n}+J_{2}^{'}b_{n}^{\dag}a_{n+1}\right],\cr\cr
H_{L_{2}}&=&\sum_{n}\left[J_{1}^{'}B_{n}^{\dag}A_{n}+J_{1}A_{n}^{\dag}B_{n}+J_{2}^{'}A_{n+1}^{\dag}B_{n}+J_{2}B_{n}^{\dag}A_{n+1}\right],\cr\cr
H_{\mathrm{Link}}&=&J_{2}Q^{\dag}b_{N}+J_{2}^{'}b_{N}^{\dag}Q+J_{2}^{'}A_{1}^{\dag}Q+J_{2}Q^{\dag}A_{1}.
\end{eqnarray}
\end{widetext}
where $a_{n}^{\dag}$ ($a_{n}$), $A_{n}^{\dag}$ ($A_{n}$), and $Q^{\dag}$ ($Q$) are the creation (annihilation) operators of the resonators $a_{n}$, $A_{n}$, and $Q$, respectively. $H_{L_{1}}$ ($H_{L_{2}}$) is the Hamiltonian of the resonator chains $L_{1}$ ($L_{2}$) with the nonreciprocal couplings $J_{1}=t_{1}+\delta$, $J_{1}^{'}=t_{1}-\delta$, $J_{2}=t_{2}+\delta$, and $J_{2}^{'}=t_{2}-\delta$. $H_{\mathrm{Link}}$ represents the coupling between two resonator chains assisted via the auxiliary resonator $Q$. We stress that the nonreciprocal coupling between two adjacent resonators, such as $J_{1}$ and $J_{1}^{'}$, can be achieved via an auxiliary microring with gain in the upper half perimeter and loss in the bottom half perimeter~\cite{longhi2015non,longhi2015robust}. In this way, the photons from the resonator $a_{n}$ to resonator $b_{n}$ pass through the half perimeter microring with gain accompanying with the amplification of hopping amplitude, while the photons from the resonator $b_{n}$ to resonator $a_{n}$ pass through the half perimeter microring with loss accompanying with the deamplification of hopping amplitude. Essentially, the amplification and the deamplification of the hopping amplitude are induced by the the synthetic imaginary gauge field $\lambda$ originating from the gain and loss of the auxiliary microring, with $Je^{\lambda}b_{n}^{\dag}a_{n}$ and $Je^{-\lambda}a_{n}^{\dag}b_{n}$. Thus, we can construct the required nonreciprocal coupling configuration via designing the appropriate synthetic imaginary gauge field $\lambda$, with $Je^{\lambda}=J_{1}$ and $Je^{-\lambda}=J_{1}^{'}$.

Obviously, when the Hamiltonian $H_{\mathrm{Link}}$ is vanishing, the present micro-resonator array is divided into three independent components with the isolated micro-resonator $Q$, the micro-resonator chain $L_{1}$, and the micro-resonator chain $L_{2}$. Note that the independent micro-resonator chains $L_{1}$ and $L_{2}$ are the analogous version mentioned in Ref.~\cite{yokomizo2019non}, in which the chain $L_{1}$ ($L_{2}$) exhibit a non-Hermitian skin effect towards the edge resonator $b_{N}$ ($A_{1}$) since the intra-cell and the inter-cell couplings satisfy $J_{1}>J_{1}^{'}$ and $J_{2}>J_{2}^{'}$. However, when the Hamiltonian $H_{\mathrm{Link}}$ is added into the system, the isolated micro-resonator $Q$ tends to become an ensemble with the micro-resonator chains $L_{1}$ and $L_{2}$, leading the original micro-resonator chains $L_{1}$ and $L_{2}$ be equivalent to the two new odd-size micro-resonator chains $L_{1}^{'}$ and $L_{2}^{'}$. 

Thus, the two new chains $L_{1}^{'}$ and $L_{2}^{'}$ generate a interface at the auxiliary resonator $Q$. We note that, in Ref.~\cite{ezawa2019non}, an analogous nonreciprocal single energy band model with the interface has been detailed investigated. They have demonstrated that, depending on the different nonreciprocal coupling configurations, the interface state can be analytically solved accompanying with two sets of solutions, in which one solution reveals that the interface state is localized near the interface with an exponential amplification distributions towards the interface site, while another set of solution indicates that the interface state can be bounded at the two lateral sites around the interface site or be bounded at the interface site. Similar with the results obtained in Ref.~\cite{ezawa2019non}, we find that the Hamiltonian of our system, under the open boundary condition, can also be exactly diagonalized with the only one of analytical zero energy mode. And the corresponding eigenstate $|\Psi\rangle_{E=0}$, in the basis of $|0\rangle_{a_{1}}\otimes|0\rangle_{b_{1}}\otimes...\otimes|0\rangle_{a_{N}}\otimes|0\rangle_{b_{N}}\otimes|0\rangle_{Q}\otimes|0\rangle_{A_{1}}\otimes|0\rangle_{B_{1}}\otimes...\otimes|0\rangle_{A_{N}}\otimes|0\rangle_{B_{N}}$, can be analytically written as
\begin{widetext}
\begin{eqnarray}\label{e02}
|\Psi\rangle_{E=0}=|1,~0,~-\frac{t_{1}+\delta}{t_{2}-\delta},~...,~(-\frac{t_{1}+\delta}{t_{2}-\delta})^{N},~0,~(-\frac{t_{1}+\delta}{t_{2}-\delta})^{N+1},~0,~(-\frac{t_{1}+\delta}{t_{2}-\delta})^{N},~...,~-\frac{t_{1}+\delta}{t_{2}-\delta},~~0,~1\rangle.
\end{eqnarray}          
\end{widetext}
Obviously, the eigenstate $|\Psi\rangle_{E=0}$ is the interface state, in which the interface state has the maximal distribution at the resonator $Q$ when $|\frac{t_{1}+\delta}{t_{2}-\delta}|>1$, accompanying with the exponential decay distributions at the $a$-type ($B$-type) resonators towards the resonators $a_{1}$ and $B_{N}$. Especially, when $J_{1}^{'}=J_{2}^{'}=0$~\cite{ezawa2019non}, we find that all the eigenvalues of the system are zero only corresponding to two nontrivial (nonzero) eigenstates, with $|\Psi\rangle^{(1)}_{E=0}=|0,~0,~...,~0,~1,~0,~1,~0,~...,~0,~0\rangle$ and $|\Psi\rangle^{(2)}_{E=0}=|0,~0,~...,~0,~0,~1,~0,~0,~...,~0,~0\rangle$. It means that the interface state can also be localized at the resonators $b_{N}$ and $A_{1}$ with a bound state or be localized at the interface resonator $Q$ with a bound state. We think that the bound interface state $|\Psi\rangle^{(2)}_{E=0}$ essentially originates from the isolated resonator $Q$ when $H_{\mathrm{Link}}=0$. It is easy to demonstrate that the system only has one zero energy mode with its eigenstate $|0,~0,~...,~0,~0,~1,~0,~0,~...,~0,~0\rangle$ when $H_{\mathrm{Link}}=0$. In this way, when $H_{\mathrm{Link}}$ is added into the system, $H_{\mathrm{Link}}$ leads the right (left) edge state of the original chain $L_{1}$ ($L_{2}$) to be bounded. Stated thus, the zero energy interface states of the system mainly have three distributions, namely, with exponential amplification distributions towards the resonator $Q$, with the bound state being localized at resonators $b_{N}$ and $A_{1}$, and with the bound state being localized at the resonator $Q$. And we predict that, similar with the non-Hermitian skin effect, the eigenstates in our system also exhibit a skin effect towards the interface state.  

To further clarify the insightful physics, we simulate the energy spectra of the micro-resonator array versus the parameter $t_{2}$ numerically when $H_{\mathrm{Link}}$ is added into the system, as shown in Figs.~\ref{fig2}(a) and~\ref{fig2}(b). For simplicity, we focus on the case of $t_{2}>0$ in the following discussions.
We find that, when $t_{2}$ approximately satisfies $0<t_{2}<0.6$, the real energy spectrum of the system has three degenerate zero energy modes locating in the energy gap, while the three degenerate zero energy modes transform into one zero energy mode when $t_{2}$ approximately satisfies $0.6<t_{2}<1$. Especially, when $t_{2}$ is large enough with $t_{2}>1$, the two energy bands gradually touch each other, and the zero energy mode integrates into the bulk states at the same time. Note that the imaginary part of the energy spectrum disappears corresponding to a pure real energy spectrum when $t_{2}$ approximately satisfies $t_{2}>0.8$. Besides, as the predication mentioned above, we find that all the eigenstates of the system indeed exhibit a localized effect, which is similar with the non-Hermitian skin effect~\cite{yokomizo2019non}, as shown in Figs.~\ref{fig2}(c) and~\ref{fig2}(d).   

Together with the analytical solutions of the three zero energy interface states, the reason for the existence of the three degenerate zero energy modes in energy spectrum can be further interpreted as follows. When the coupling parameter satisfies $0<t_{2}<0.6$, it corresponds that a relatively mild $H_{\mathrm{Link}}$ is added into the system, leading the original uncorrelated three sub-components $Q$, $L_{1}$, and $L_{2}$ generate the following inclinations: $(\mathrm{\rmnum{1}})$ the original isolated auxiliary resonator $Q$ remains to keep the isolated inclination accompanying with the generation of the bound state; $(\mathrm{\rmnum{2}})$ the two resonator chains $L_{1}$ and $L_{2}$ remain to keep the independent inclinations accompanying with the original right edge resonator $b_{N}$ of $L_{1}$ and the original left edge resonator $A_{1}$ of $L_{2}$ (Note that the equivalent inclination is that the original right edge resonator $b_{N}$ of $L_{1}$ and the original left edge resonator $A_{1}$ of $L_{2}$ are bounded due to $H_{\mathrm{Link}}$); $(\mathrm{\rmnum{3}})$ the two resonator chains $L_{1}$ and $L_{2}$ also tend to link with the auxiliary resonator $Q$ accompanying with the generations of the new right edge resonator $Q$ of $L_{1}^{'}$ and the new left edge resonator $Q$ of $L_{2}^{'}$. Thus, as revealed in the analytical solutions, these three inclinations imply that the three degenerate zero energy modes correspond to the three interface states with being bounded at interface resonator $Q$, being bounded at resonators $b_{N}$ and $A_{1}$, and being localized at resonator $Q$ with the exponential distributions respectively. In other words, these inclinations originating from the weak interaction $H_{\mathrm{Link}}$ generate effects on both the isolated bound state, original topological edge states, and the new topological edge states simultaneously. Specially, we stress that the effects of the weak interaction $H_{\mathrm{Link}}$ on the original topological edge states and the new topological edge states will directly determine the skin effects of the original two resonator chains $L_{1}$ and $L_{2}$ and the two new resonator chains $L_{1}^{'}$ and $L_{2}^{'}$. Thus, these Joint effects will together determine the zero energy modes and the states of the system.  

\subsection{\label{sec.B}The numerical analysis for weak $t_{2}$}

To further demonstrate the points mentioned above, we plot the energy spectra and the distributions of the zero energy modes when an extremely weak $t_{2}=0$ is added into the system, as shown in Fig.~\ref{fig3}. The numerical results show that, indeed, the real and the imaginary parts of the energy spectrum both have three degenerate zero energy modes locating in the gap, in which the zero energy modes are divided into three types. More specifically, as shown in Fig.~\ref{fig3}(a), the two zero energy modes in the real energy spectrum exhibit a characteristic of the bound states, in which the photons mainly gather around the auxiliary micro-resonator $Q$. And there is also one zero energy mode being localized near the auxiliary micro-resonator $Q$ with the exponential attenuation or amplification distributions at the odd resonators, which means that the zero energy mode is a topological zero energy mode corresponding to the generations of the new right edge of $L_{1}^{'}$ and the new left edge of $L_{2}^{'}$. Note that the generations of the new topological edges leads that the two new micro-resonator chains $L_{1}^{'}$ and $L_{2}^{'}$ both exhibit a non-Hermitian skin effect towards the auxiliary micro-resonator $Q$. 

Besides, as shown in Fig.~\ref{fig3}(b), the imaginary energy spectrum of the system also possesses three degenerate zero energy modes, in which two zero energy modes are localized near the resonators $b_{N}$ and $A_{1}$ with the the approximatively exponential attenuation or amplification distributions at the even resonators (We stress that the two zero energy states are the bound states in essence because of the weak $H_{\mathrm{Link}}$. However, to interpret it more intuitive, for the extremely weak $H_{\mathrm{Link}}\sim0$, we regard these two zero energy states as the original topological edge states of $L_{1}$ and $L_{2}$). It reveals that the two zero energy modes correspond to the original topological right edge $b_{N}$ of $L_{1}$ and the original topological left edge $A_{1}$ of $L_{2}$, which means that the original micro-resonator chains $L_{1}$ ($L_{2}$) still tends to exhibit a non-Hermitian skin effect towards the original topological right (left) edge $b_{N}$ ($A_{1}$). Note that the rest one of the zero energy mode corresponds to the new topological edges of $L_{1}^{'}$ and $L_{2}^{'}$, which is the same zero energy mode discussed in the real energy spectrum. 

To further verify the above conclusions, we simulate the distributions of all the eigenstates, as shown in Figs.~\ref{fig3}(c) and~\ref{fig3}(d). The numerical results reveal that all the eigenstates are divided into two types, in which one type of eigenstates has the maximal distribution at the resonator $Q$ (labeled by $\mathrm{\Rmnum{1}}$-type eigenstates) while another type of eigenstates has the maximal distributions at the resonators $b_{N}$ and $A_{1}$ (labeled by $\mathrm{\Rmnum{2}}$-type eigenstates). Obviously, the two types of eigenstates exhibit a similar skin effect, which is consistent with the conclusions mentioned above. Thus, it indicates that all the photons will mainly gather into the resonators $b_{N}$, $Q$, and $A_{1}$ since all the eigenstates are both localized near the resonators $b_{N}$, $Q$, and $A_{1}$. The localization of all the eigenstates is caused by the similar non-Hermitian skin effects and the bound effect of the isolated resonator $Q$. At the same time, we find that the two types of eigenstates approximately have the same order of magnitude for the distributions, which means that the photons will mainly appear into the resonators $b_{N}$, $Q$, and $A_{1}$ with the same proportion.    
      
The properties that all the photons gather towards the certain micro-resonators have widely potential applications in the photon storage device~\cite{maxwell2013storage,yanik2007dynamic} and the laser generator device~\cite{charles2015selective,bandres2018topological,harari2018topological}. Thus, to further clarify it, we investigate the evolution of the photons in the resonator array if we excite the different resonators of the resonator array initially. The numerical results show that, as shown in Fig.~\ref{fig4}(a), the photons mainly gather into the resonators $b_{N}$, $Q$, and $A_{1}$ with the developing of the time when the auxiliary micro-resonator $Q$ is excited initially. The reason that all the photons are mainly gathered into the resonators $b_{N}$, $Q$, and $A_{1}$ in the approximatively uniform way is that the two types of eigenstates approximately have the same order of magnitude for the distributions. When other resonators are excited initially, the photons still mainly gather into the resonators $b_{N}$, $Q$, and $A_{1}$ with the developing of the time, as shown in Figs.~\ref{fig4}(b)-\ref{fig4}(d).

\subsection{\label{sec.C}The effects of the increasing $t_{2}$}

\begin{figure}
	\centering
	\subfigure{\includegraphics[width=0.49\linewidth]{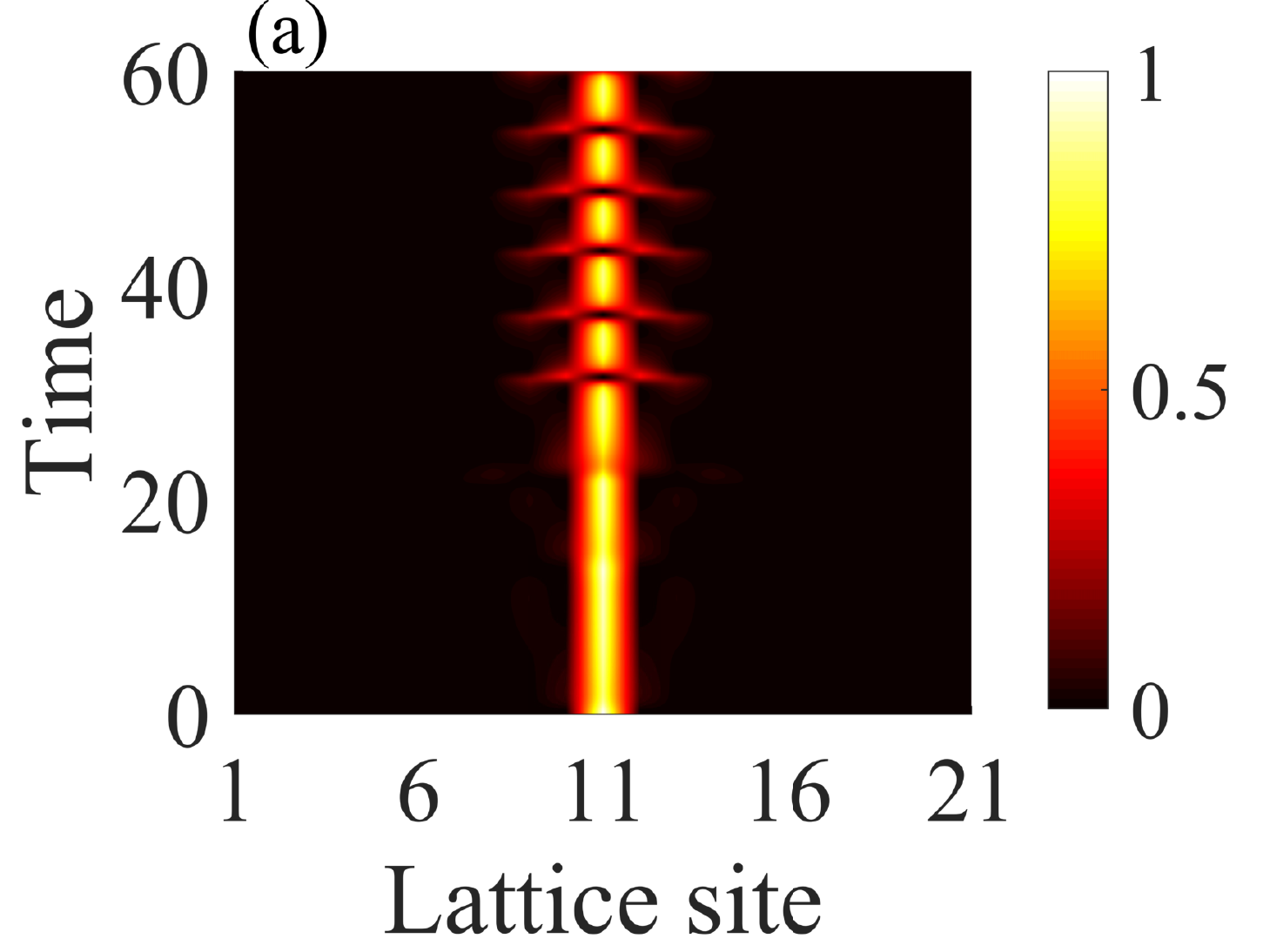}}
	\subfigure{\includegraphics[width=0.49\linewidth]{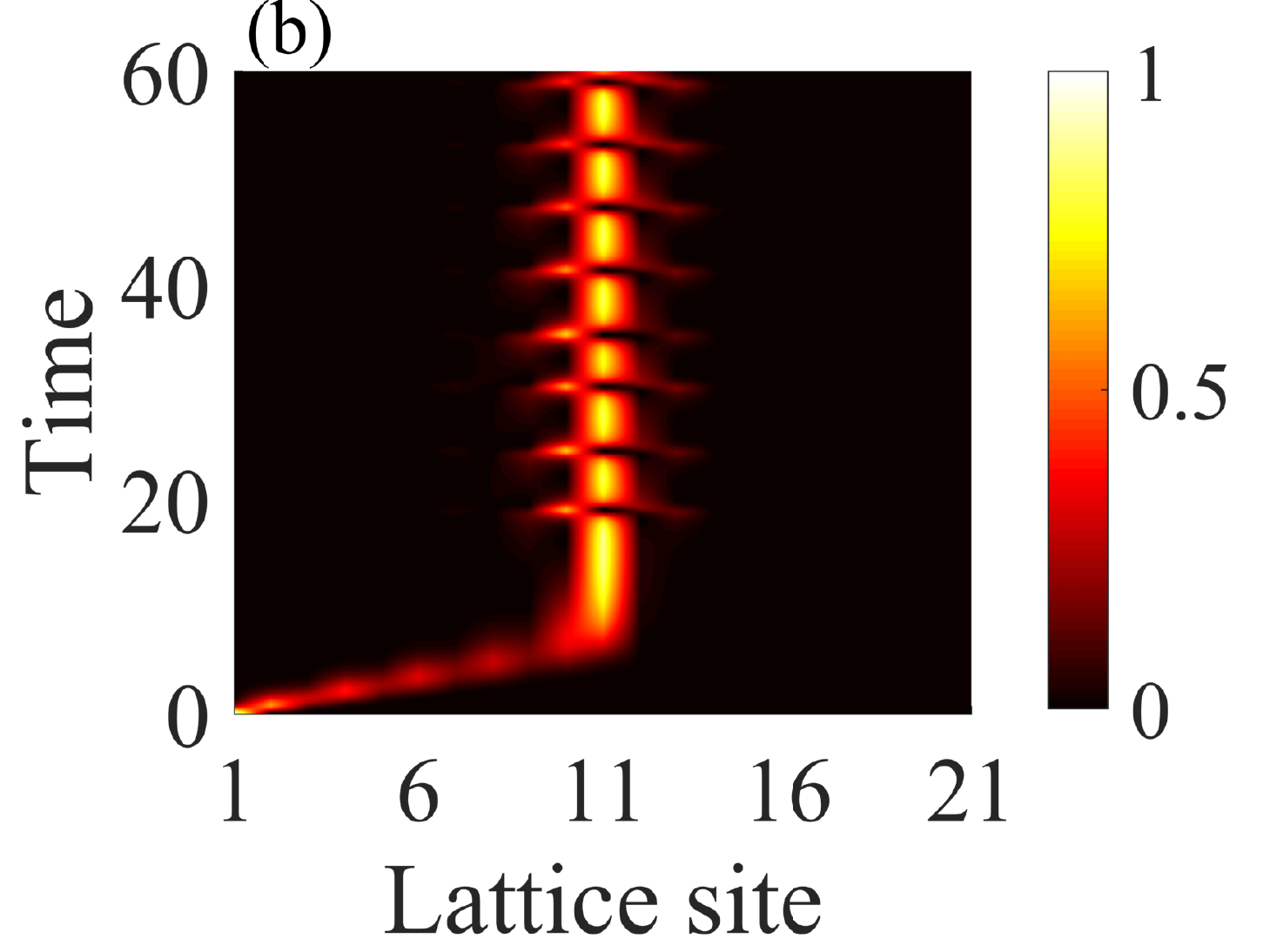}}
	
	\subfigure{\includegraphics[width=0.49\linewidth]{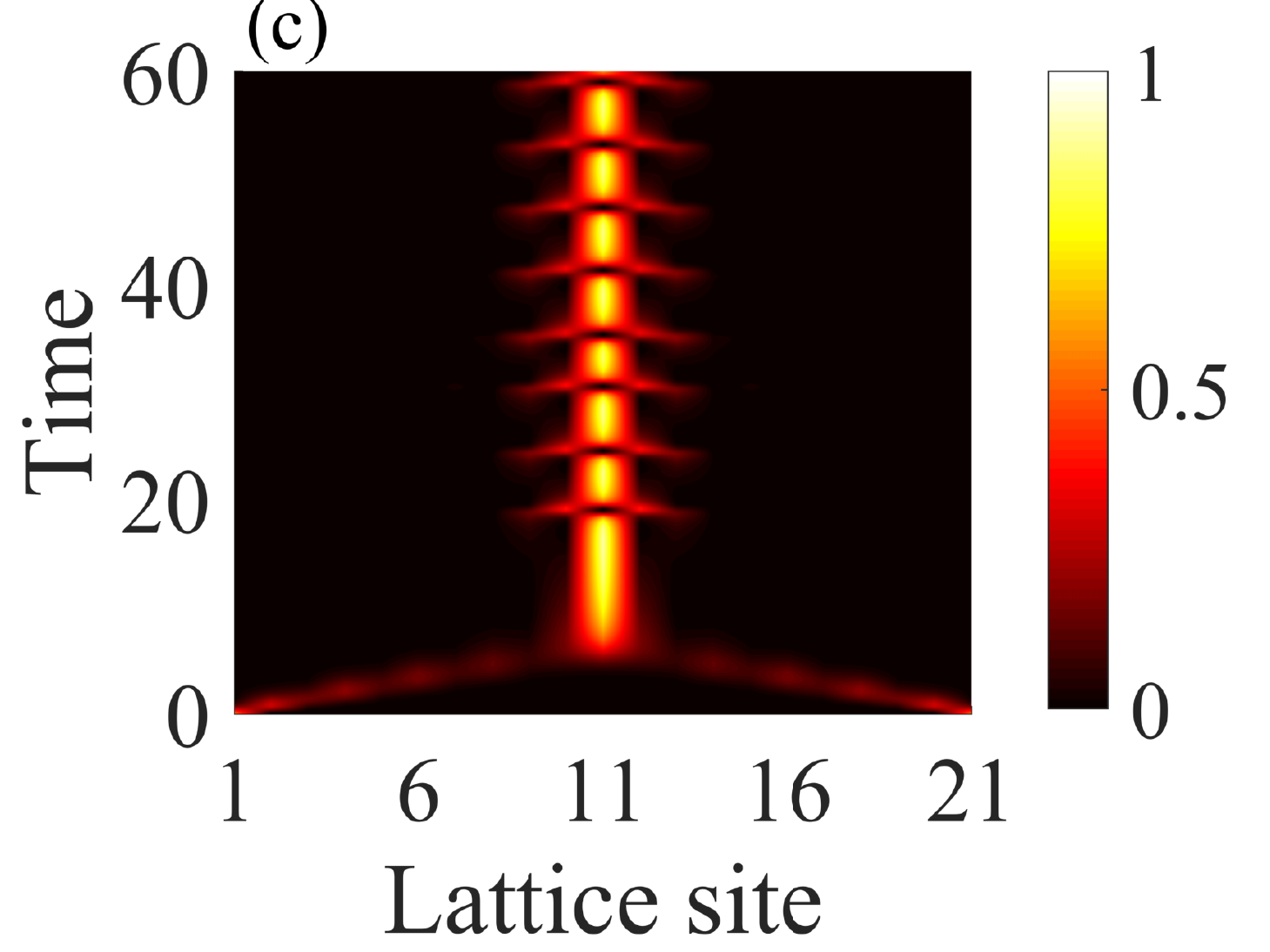}}
	\subfigure{\includegraphics[width=0.49\linewidth]{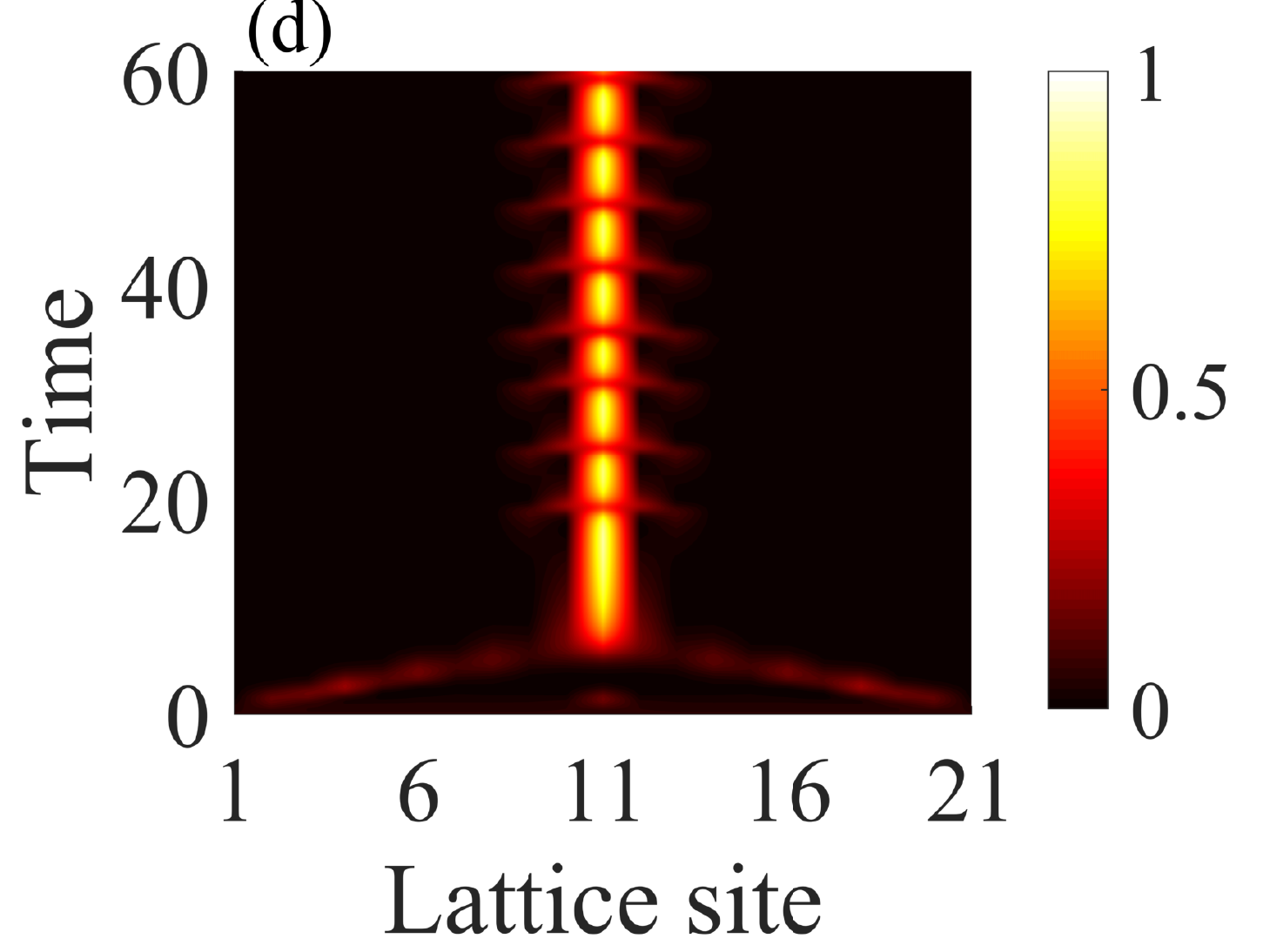}}\caption{The interface-state laser in the non-Hermitian resonator array when $L_{1}=L_{2}=10$. (a) The evolution of the photons when the auxiliary resonator $Q$ is excited. (b) The evolution of the photons when the first resonator is excited. (c) The evolution of the photons when the first and the last resonators are excited. (d) The evolution of the photons when all of the resonators are excited. Other parameters take $t_{1}=1$, $\delta=0.8$ and $t_{2}=0.5$. We set $t_{1}$ as the unit of energy.}\label{fig6}
\end{figure}

As mentioned above, we realize the gathering of the photons towards the certain resonators. However, we except that the photons mainly gather into the one certain resonator to obtain the excellent performance of photonic gathering, which is crucial to realize the laser generator device. Thus, we focus on the case that $t_{2}$ is gradually increased. For example, when $t_{2}=0.5$, we find that the real part and the imaginary part of the energy spectrum still have three degenerate zero energy modes, as shown in Fig.~\ref{fig5}. The numerical results show that, with the increasing of $t_{2}$, the bound effect of the original isolated resonator $Q$ is weakened while the new topological edge states of the new resonator chains $L_{1}^{'}$ and $L_{2}^{'}$ are strengthened, as shown in in Fig.~\ref{fig5}(a). The reason is that the correlation between the original resonator chain $L_{1}$ ($L_{2}$) and the auxiliary resonator $Q$ is strengthened due to the existence of the strong interaction $H_{\mathrm{Link}}$. The strengthened correlation between the original resonator chain $L_{1}$ ($L_{2}$) and the auxiliary resonator $Q$ also further destroys the original topological right (left) edge $b_{N}$ ($A_{1}$) of $L_{1}$ ($L_{2}$), leading that the topological effect of the original topological right (left) edge $b_{N}$ ($A_{1}$) of $L_{1}$ ($L_{2}$) tend to become the bound effect, as shown in Fig.~\ref{fig5}(b). As a result, the gathering of the photons towards the middle resonator $Q$ is strengthened while the gathering of the photons towards the resonators $b_{N}$ and $A_{1}$ is weakened. To further demonstrate the conclusions mentioned above, we simulate the evolutions of the photons with the developing of the time when the different resonators are excited initially. The numerical results show that, as shown in Fig.~\ref{fig6}(a), the photons mainly gather into the middle resonator $Q$ and the resonators $b_{N}$ and $A_{1}$ alternatively with the developing of the time when the resonator $Q$ is excited initially. It indicates that we can realize the pulsed interface laser since the photons mainly gather into the resonator $Q$ at the interface intermittently. The numerical results when other resonators are excited initially are shown in Figs.~\ref{fig6}(b)-\ref{fig6}(d). The numerical results reveal the same conclusions discussed above.      

\section{\label{sec.3}The interface laser in the non-Hermitian topologically trivial resonator array}
\begin{figure}
	\centering
	\subfigure{\includegraphics[width=0.49\linewidth]{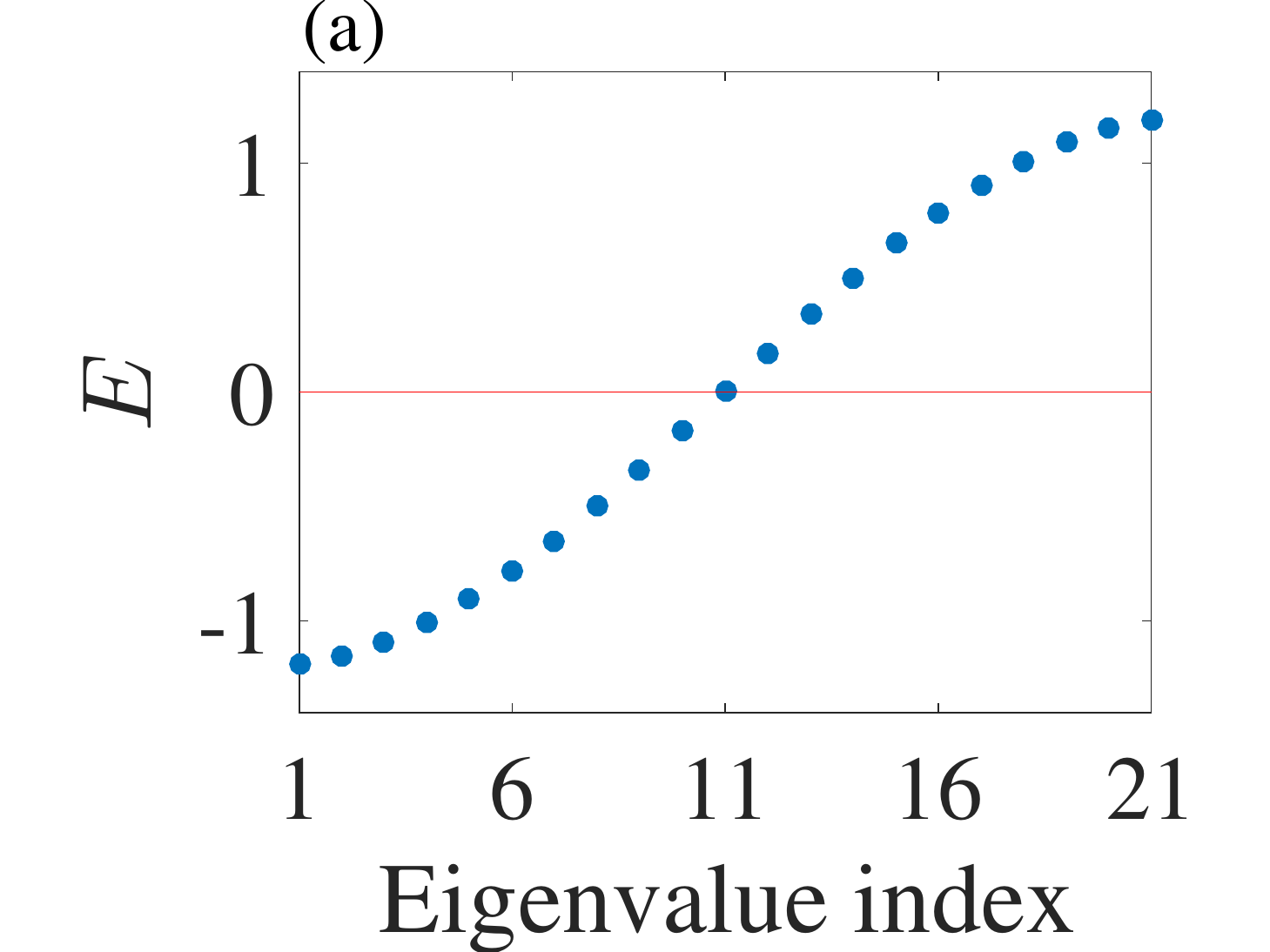}}
	\subfigure{\includegraphics[width=0.49\linewidth]{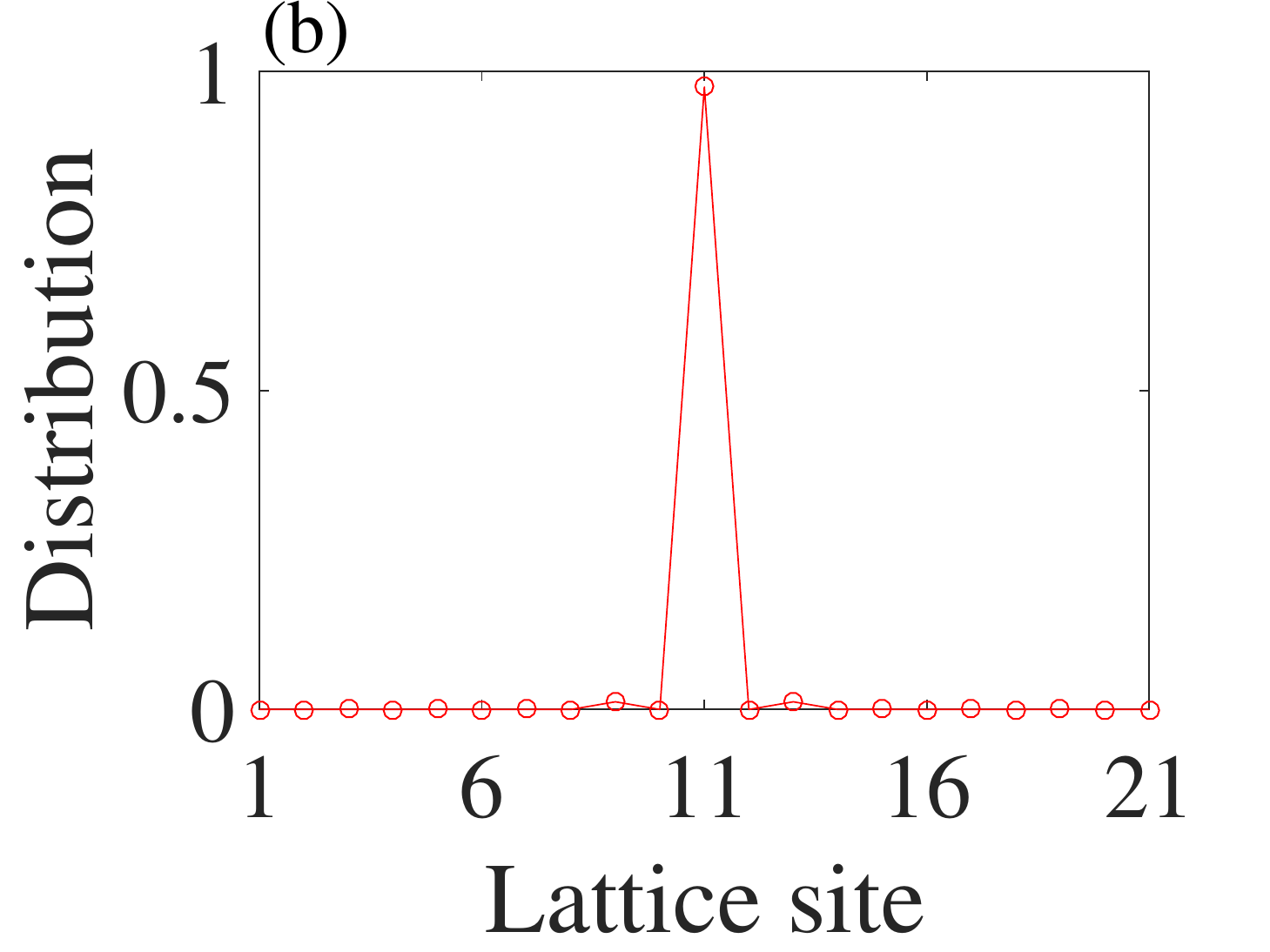}}
	
	\subfigure{\includegraphics[width=0.49\linewidth]{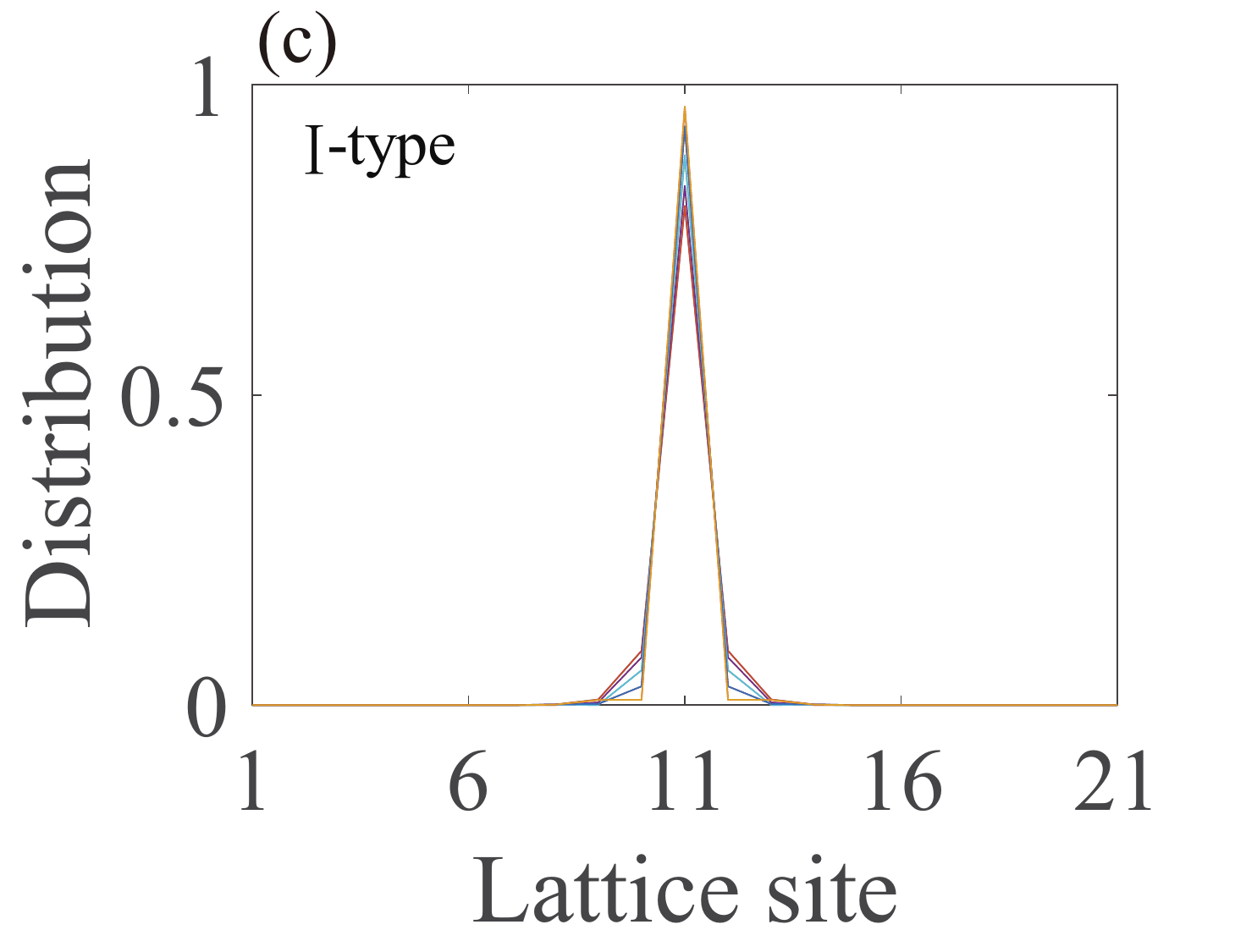}}
	\subfigure{\includegraphics[width=0.49\linewidth]{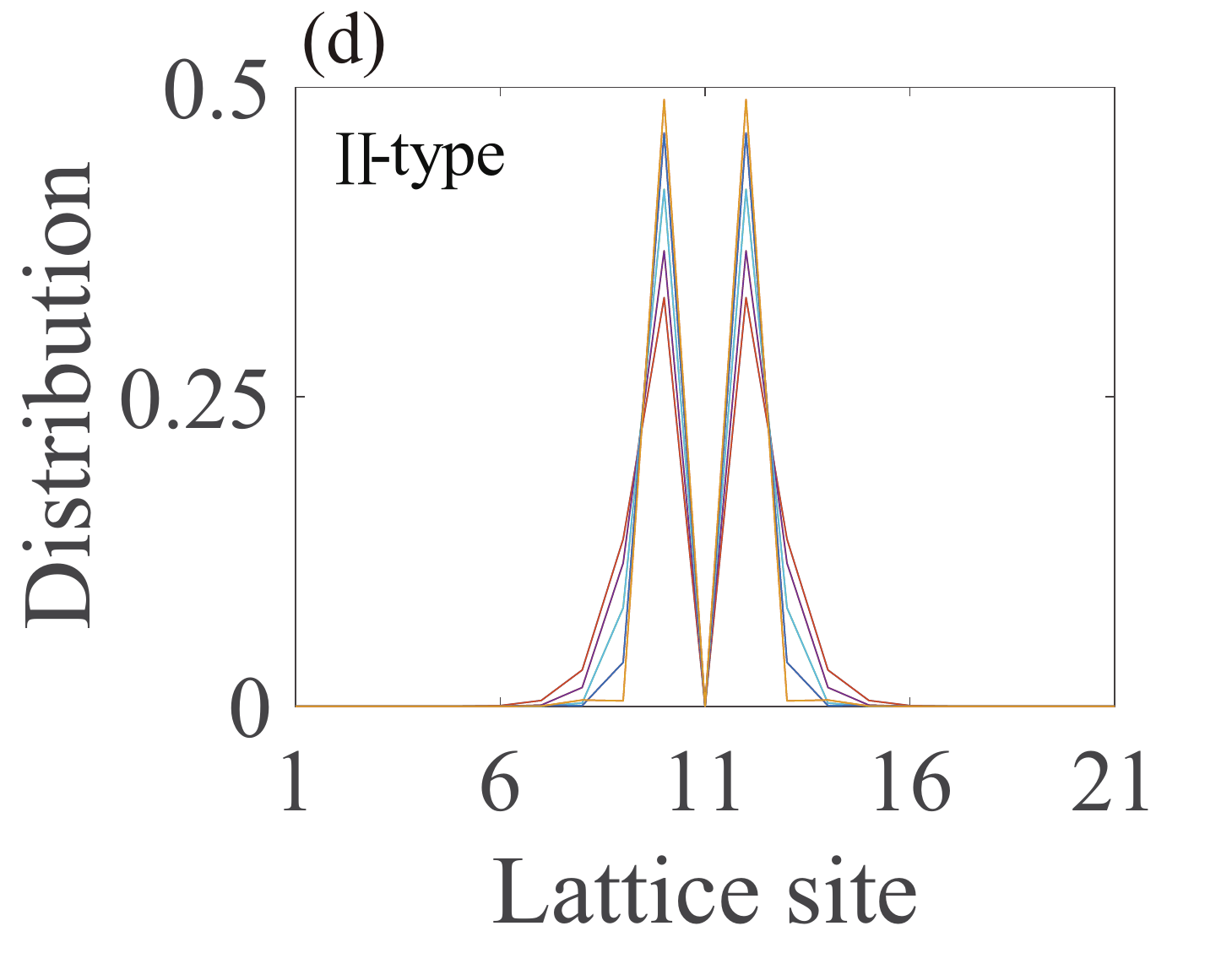}}		
	\caption{The energy spectrum and the distributions of the states when $L_{1}=L_{2}=10$. (a) The non-gapped energy spectrum of the resonator array, which still contains a zero energy mode. (b) The distribution of the zero energy mode. (c) There are $10$ eigenstates possessing the maximal distribution at the $11$th resonator $Q$. (d) There are $10$ eigenstates possessing the maximal distribution at the $10$th and the $12$th resonators. Other parameters take $t_{1}=1$, $\delta=0.8$, and $t_{2}=1$. We set $t_{1}$ as the unit of energy.}\label{fig7}
\end{figure}

Now, we further increase the strength of $t_{2}$ with $t_{2}=1$. As discussed in the last section, when $t_{2}>\delta$, we find that the system corresponds to a pure real energy spectrum. As well known, the physical meaning of the pure real energy spectrum can be revealed easier in experiment. Besides, the zero energy mode gradually integrates into the bulk state corresponding to a non-gapped energy spectrum when $t_{2}>1$. Thus, when $t_{2}=1$, the present nonreciprocal resonator array is a topologically trivial system since the resonator array has the same intra-cell and the inter-cell coupling configurations. To further clarify the topology of the system, we plot the energy spectrum of the resonator array, as shown in Fig.~\ref{fig7}(a). We find that the energy spectrum of the system does not possess a energy gap corresponding to a continuous energy spectrum. Although the system does not have the energy gap, the present system still owns a zero energy mode. Analogously, as shown in Fig.~\ref{fig7}(b), the zero energy mode has the maximal distribution at the middle resonator $Q$, which can be potentially used to realize the perfect pulsed interface-state laser. The reason of the above phenomenon is that the bound states and the original topological right (left) edge state of $L_{1}$ ($L_{2}$) are further inhibited (the bound state of the isolated $Q$ disappears and the topological edges of $L_{1}$ and $L_{2}$ disappear) due to the existence of the strong interaction $H_{\mathrm{Link}}$. As a result, the gathering of the photons into the resonator $Q$ is strengthen again accompanying with the further weakened gathering of the photons into the resonators $b_{N}$ and $A_{1}$. 

To demonstrate the above analysis, we plot the distributions of the bulk eigenstates, as shown in Figs.~\ref{fig7}(c) and \ref{fig7}(d). Similar with the case of weak $t_{2}$, we find that the bulk eigenstates of the system are also divided into two types, in which one type of the eigenstates has the maximal distribution at the middle resonator $Q$ while other eigenstates have the maximal distributions at the resonators $b_{N}$ and $A_{1}$. Obviously, the $\mathrm{\Rmnum{1}}$-type bulk states are mainly caused by the skin effect of the new resonator chains $L_{1}^{'}$ and $L_{2}^{'}$, while, the $\mathrm{\Rmnum{2}}$-type bulk states mainly originate from the bound effect after the original edges of $L_{1}$ and $L_{2}$ are further destroyed. Note that, different from the case of weak $t_{2}$, we find that the order of magnitude for the $\mathrm{\Rmnum{1}}$-type eigenstates distributions is much large than the order of magnitude for the $\mathrm{\Rmnum{2}}$-type eigenstates distributions, leading that all the photons to be easier gathered into the resonator $Q$ compared with the resonators $b_{N}$ and $A_{1}$. In this way, the photons mainly gather into the resonator $Q$ for a long time while the photons transiently gather into the resonators $b_{N}$ and $A_{1}$. Thus, we can realize the better gathering of the photons into the resonator $Q$.      
\begin{figure}
	\centering
	\subfigure{\includegraphics[width=0.49\linewidth]{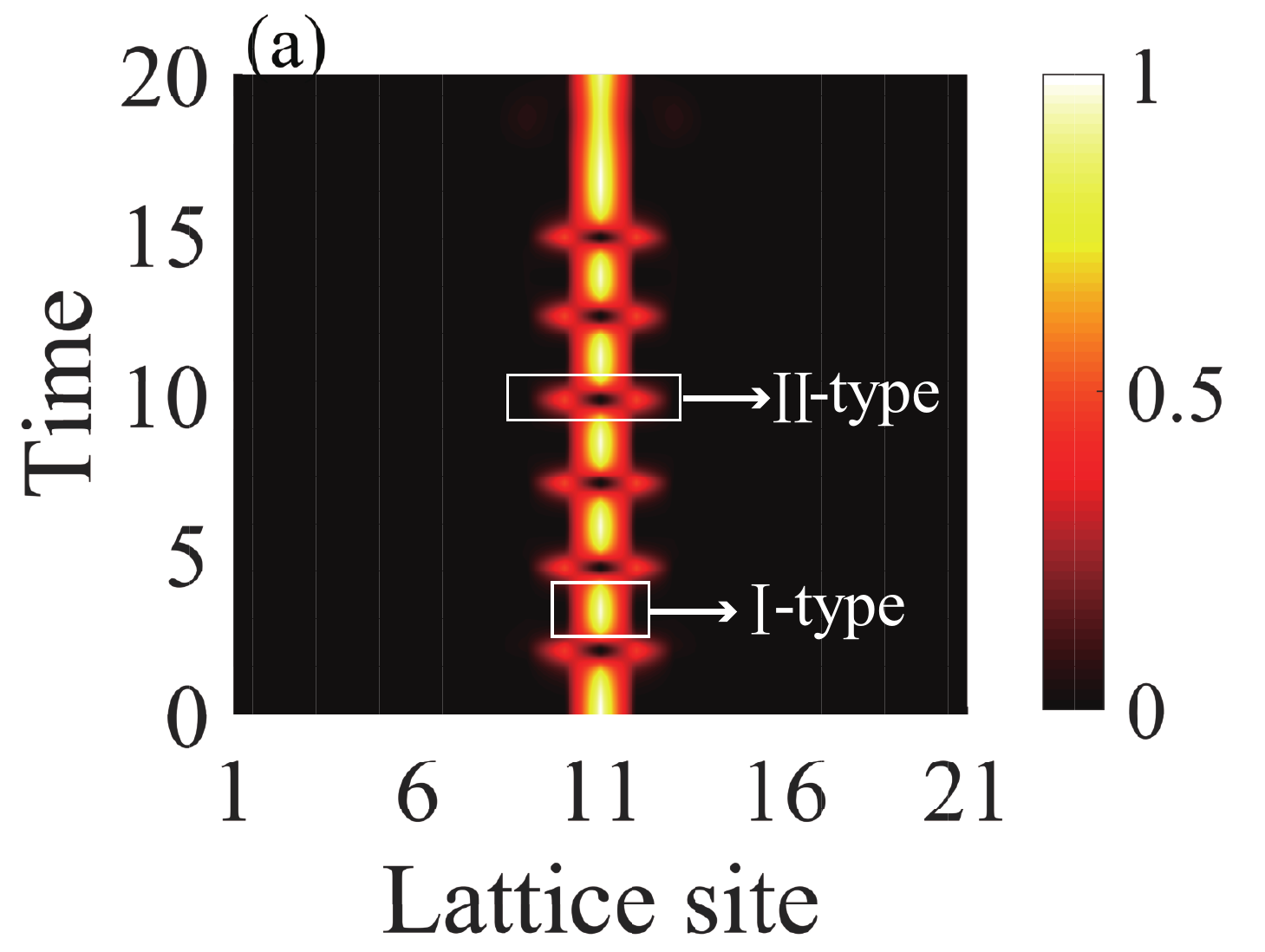}}
	\subfigure{\includegraphics[width=0.49\linewidth]{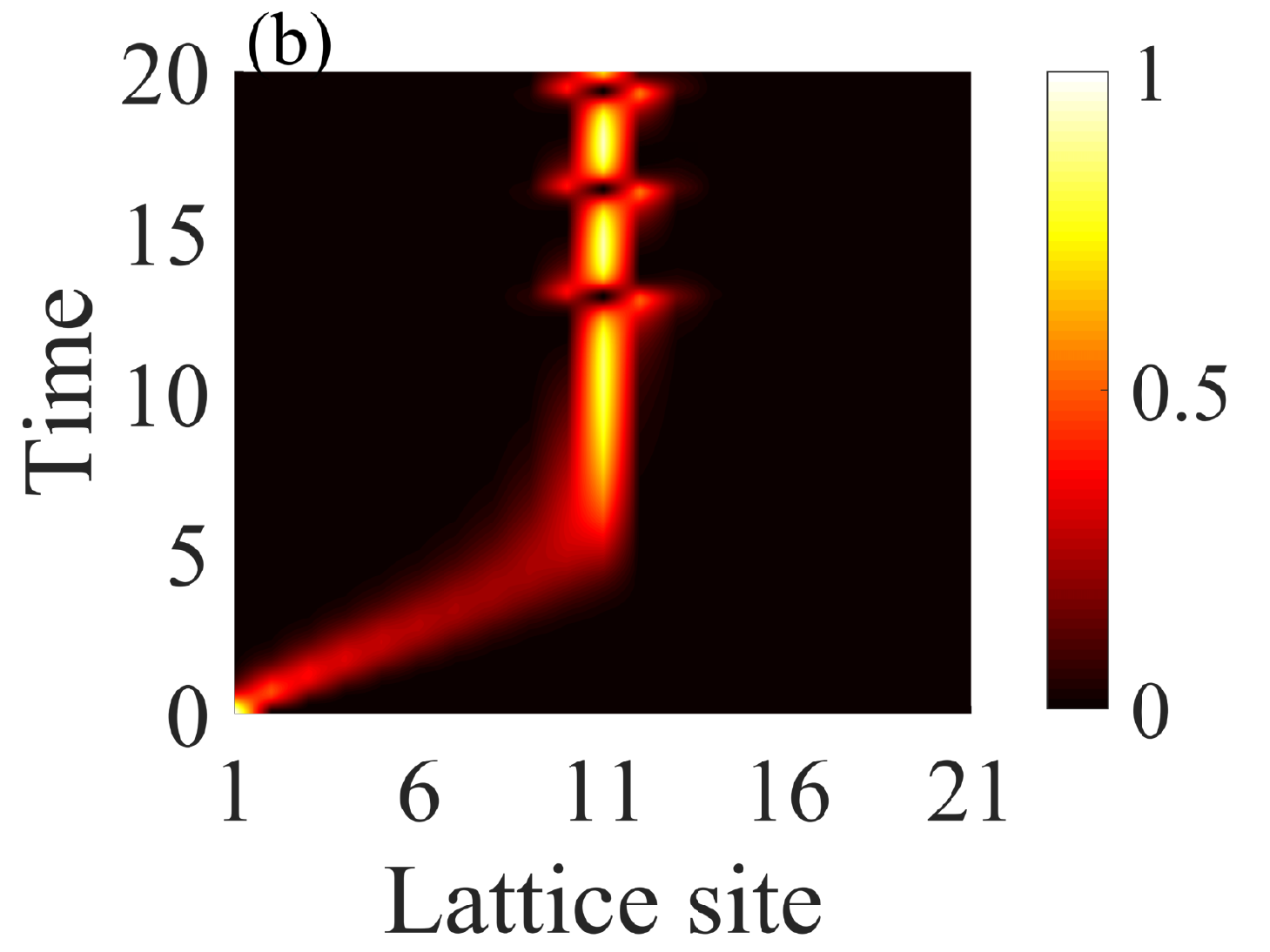}}
	
	\subfigure{\includegraphics[width=0.49\linewidth]{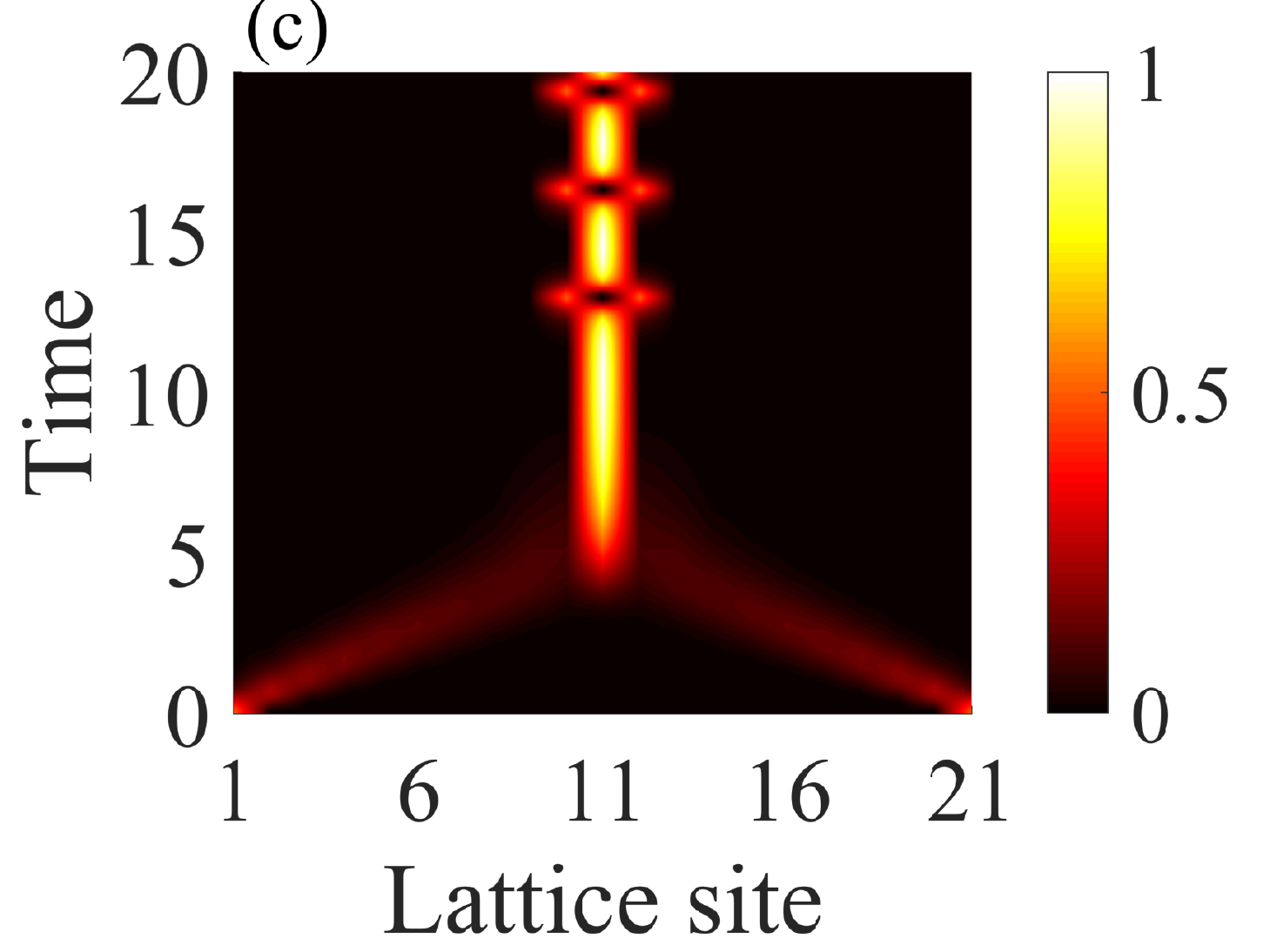}}
	\subfigure{\includegraphics[width=0.49\linewidth]{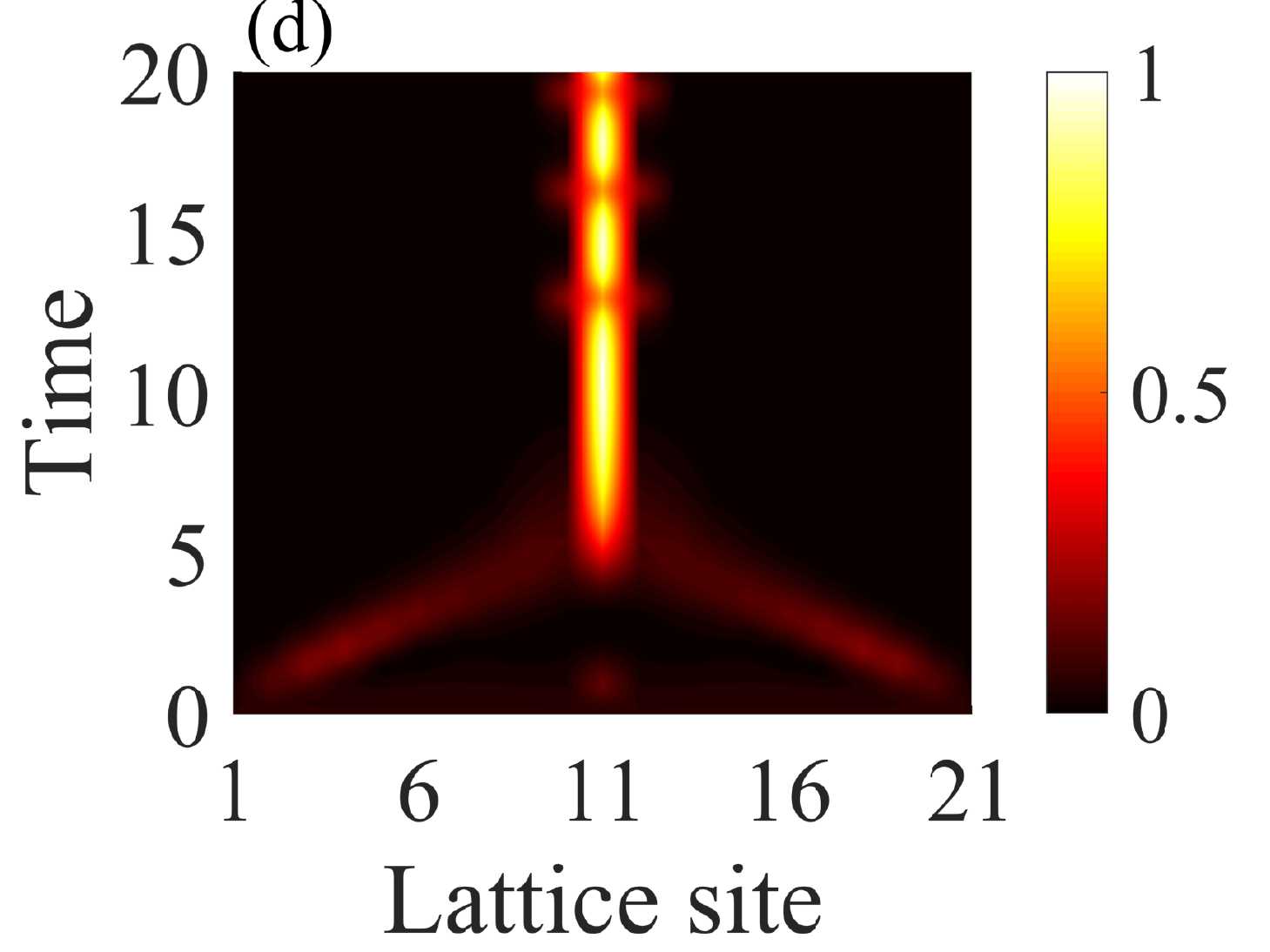}}		
	\caption{The interface-state laser in the non-Hermitian resonator array when $L_{1}=L_{2}=10$. (a) The evolution of the photons when the auxiliary resonator $Q$ is excited. (b) The evolution of the photons when the first resonator is excited. (c) The evolution of the photons when the first and the last resonators are excited. (d) The evolution of the photons when all of the resonators are excited. Other parameters take $t_{1}=1$, $\delta=0.8$, and $t_{2}=1$. We set $t_{1}$ as the unit of energy.}\label{fig8}
\end{figure} 
\begin{figure}
	\centering
	\subfigure{\includegraphics[width=0.49\linewidth]{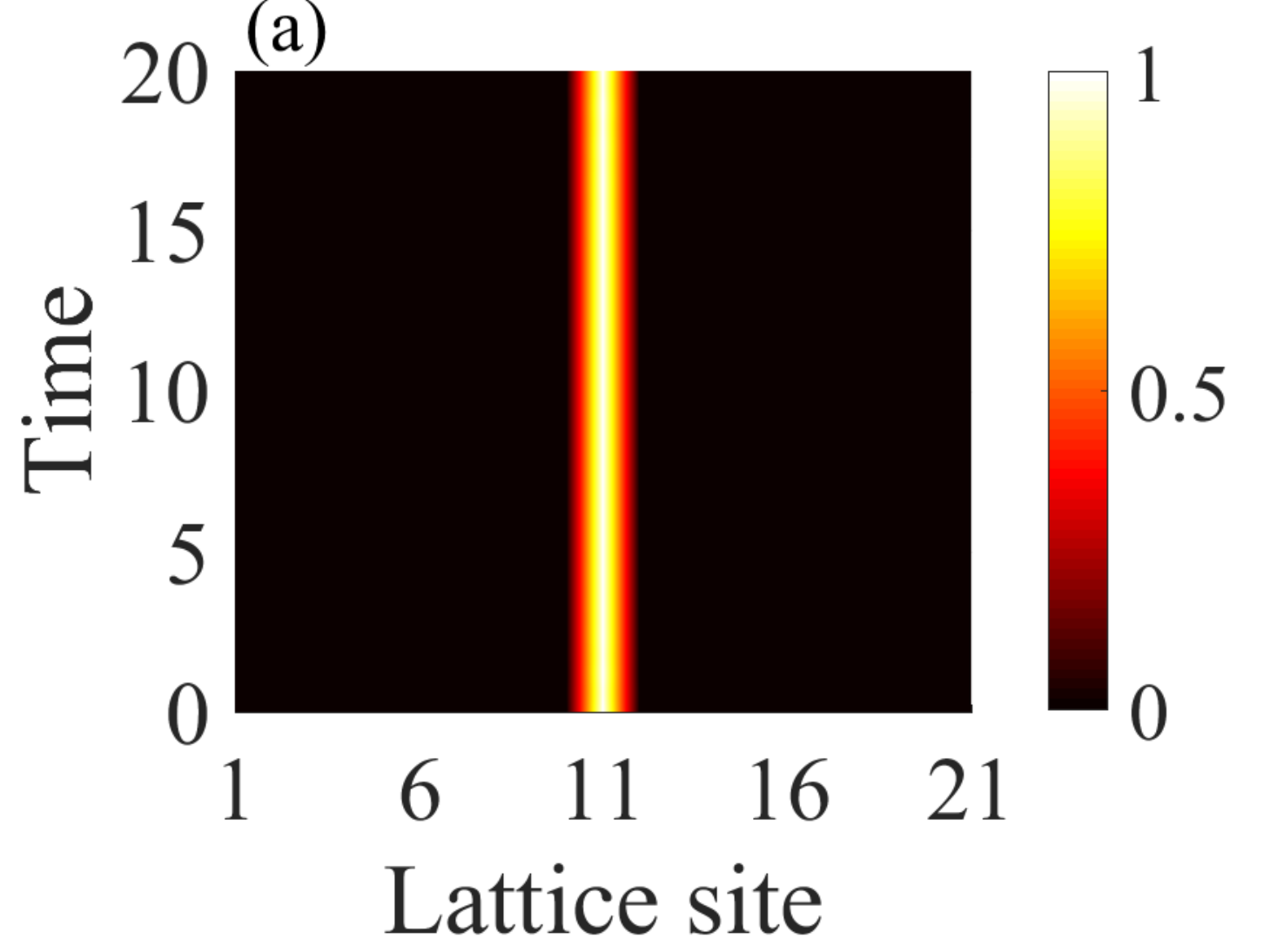}}
	\subfigure{\includegraphics[width=0.49\linewidth]{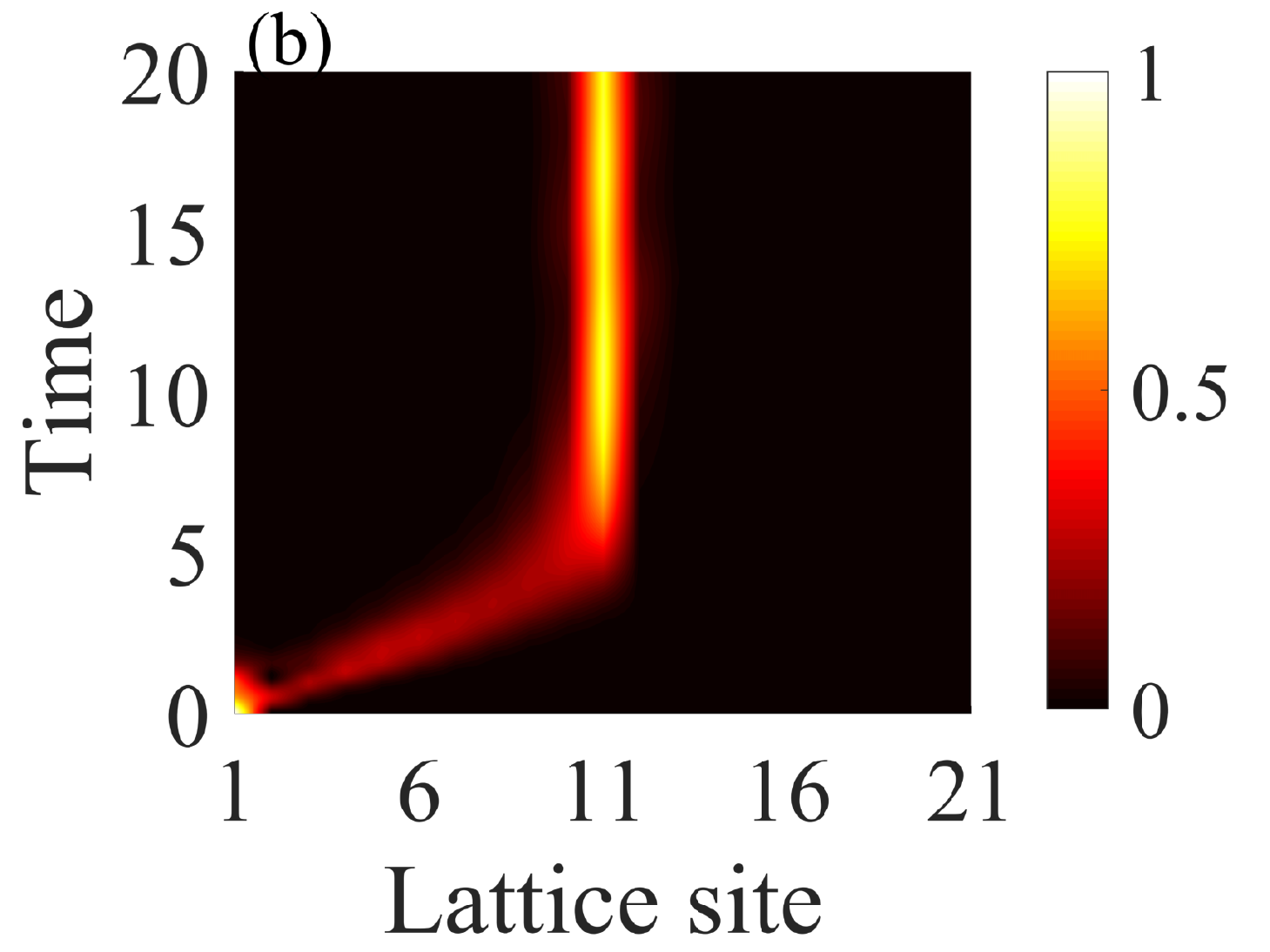}}
	
	\subfigure{\includegraphics[width=0.49\linewidth]{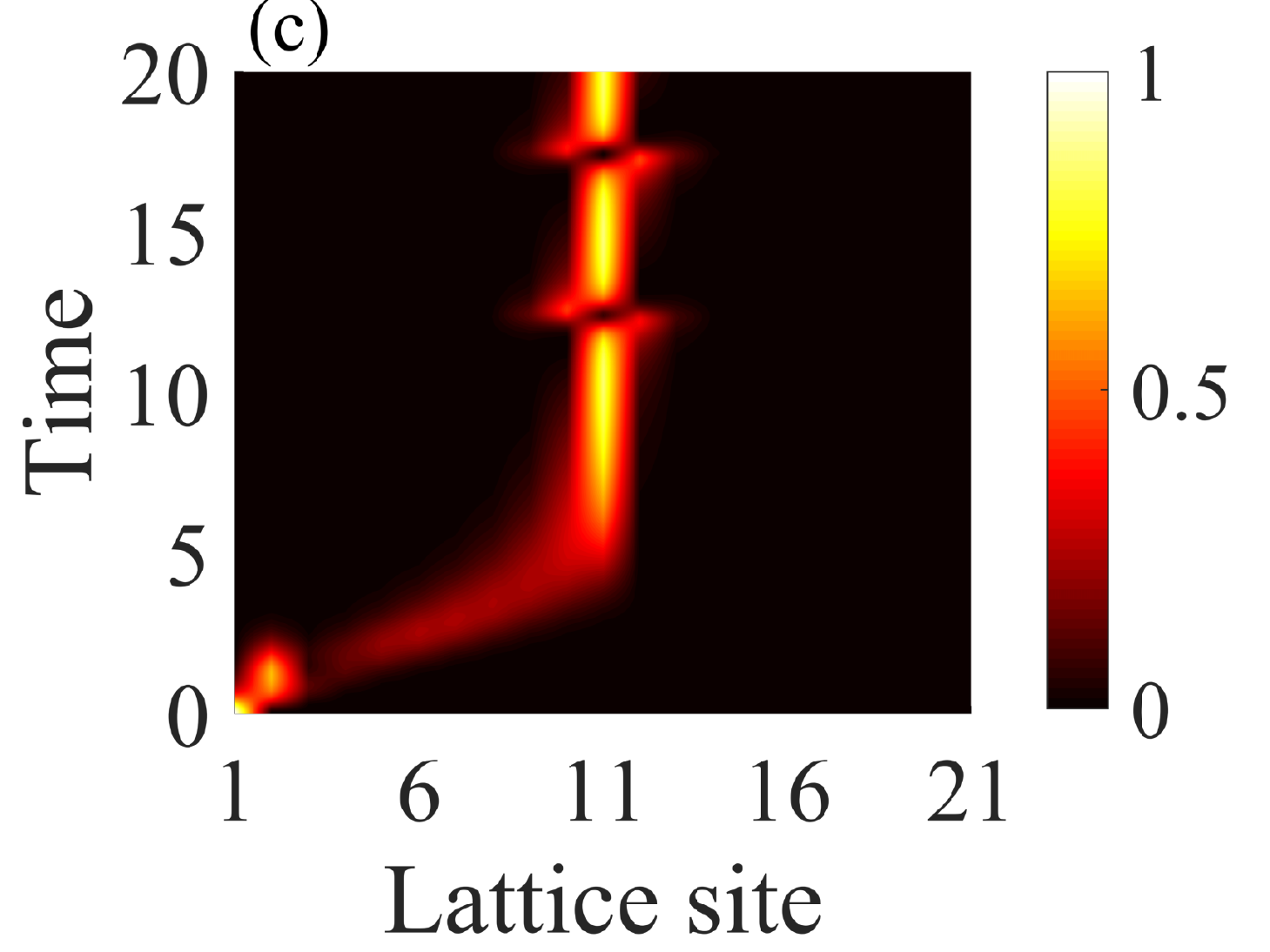}}	
	\subfigure{\includegraphics[width=0.49\linewidth]{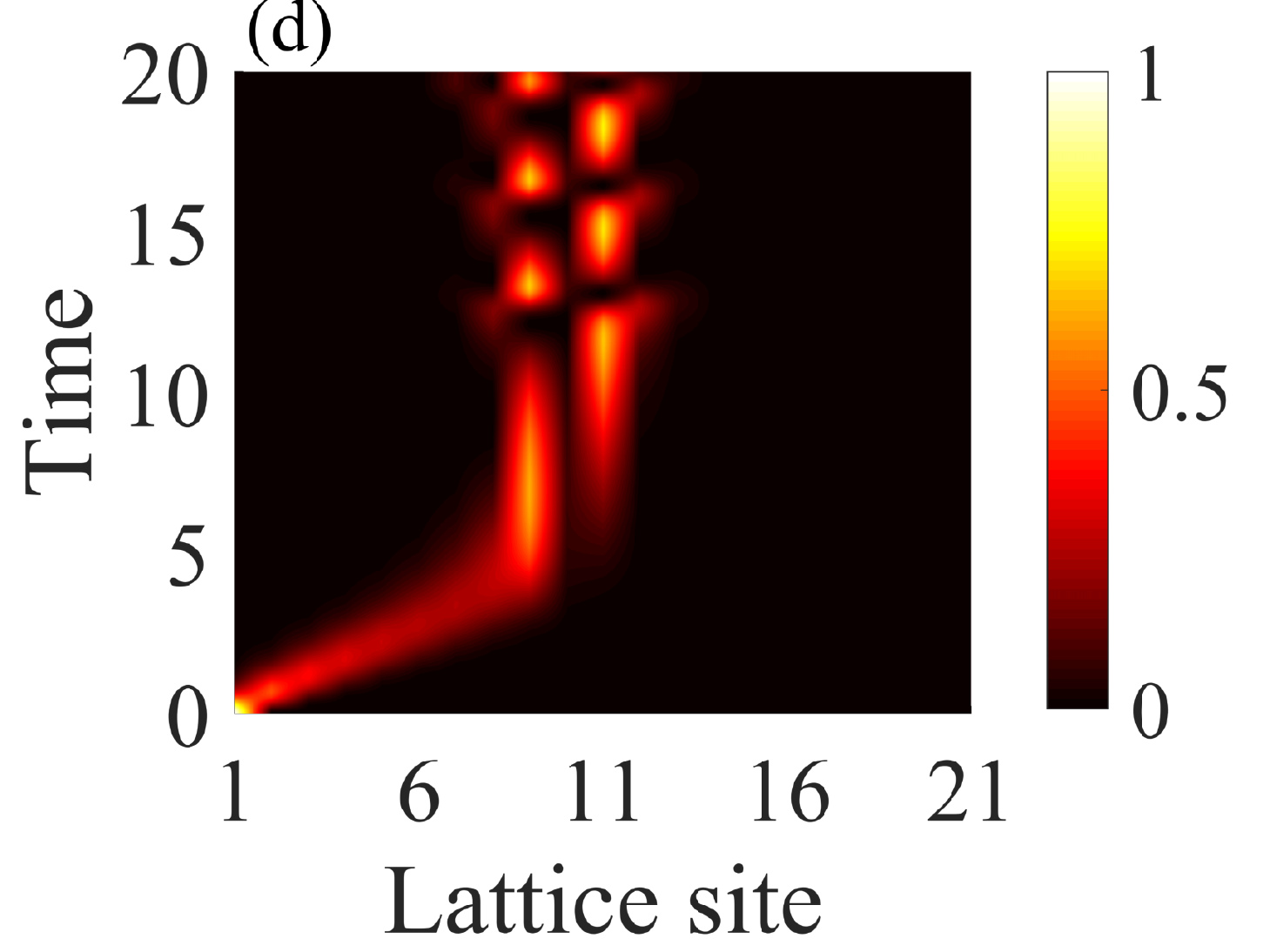}}	
	\caption{The effects of the on-site defects on the interface-state laser when $L_{1}=L_{2}=10$. (a) The effects of the on-site defect added on the auxiliary resonator $Q$ on the interface-state laser. (b) The effects of the on-site defect added on the second resonator on the evolution of the photons when the first resonator is excited. (c) The effects of the on-site defect added on the third resonator on the evolution of the photons when the first resonator is excited. (d) The effects of the on-site defect added on the $10$th resonator on the evolution of the photons when the first resonator is excited. Other parameters take $t_{1}=1$, $\delta=0.8$, and $t_{2}=1$. We set $t_{1}$ as the unit of energy.}\label{fig9}
\end{figure} 
\begin{figure}
	\centering
	\subfigure{\includegraphics[width=0.49\linewidth]{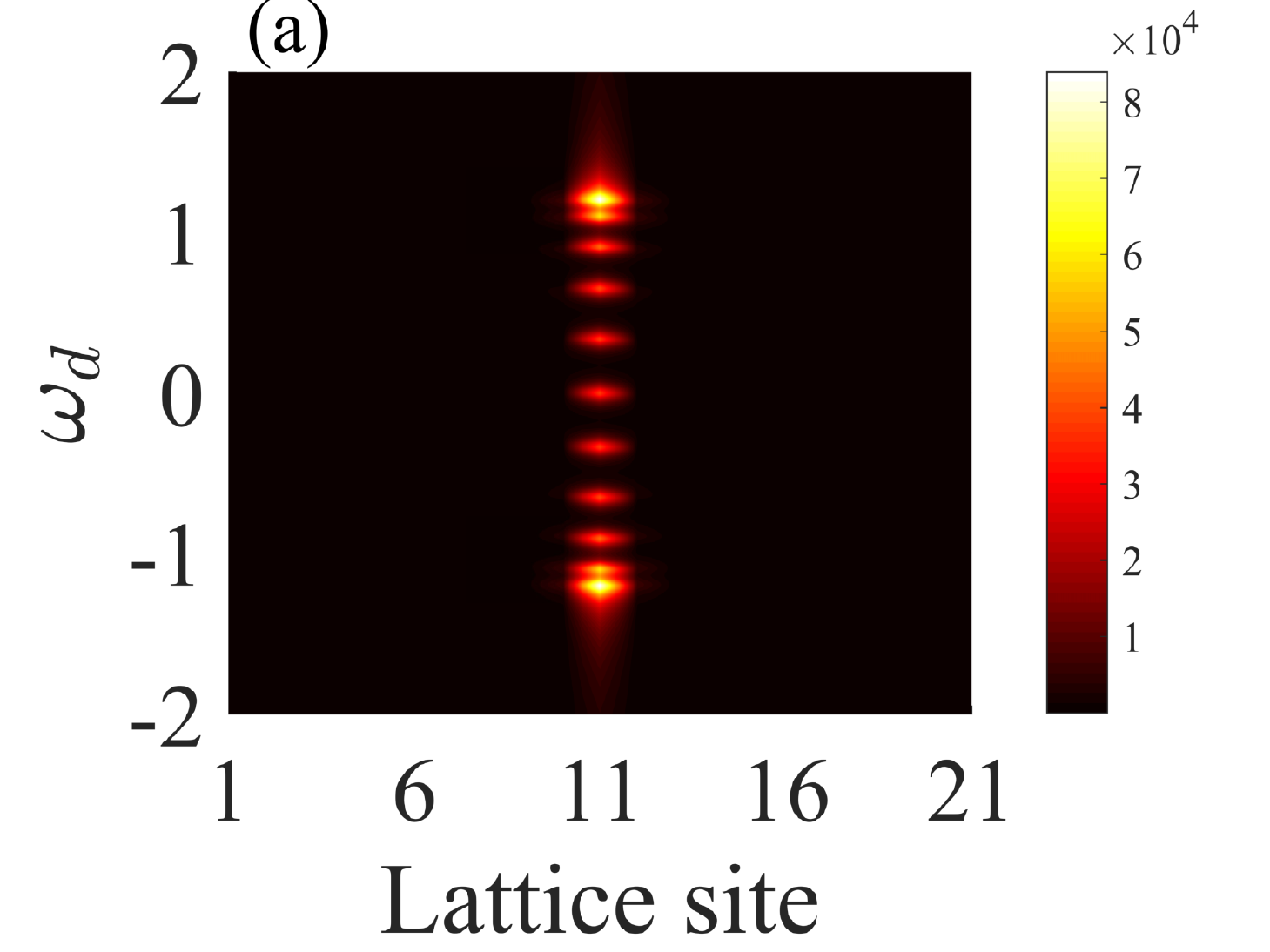}}
	\subfigure{\includegraphics[width=0.49\linewidth]{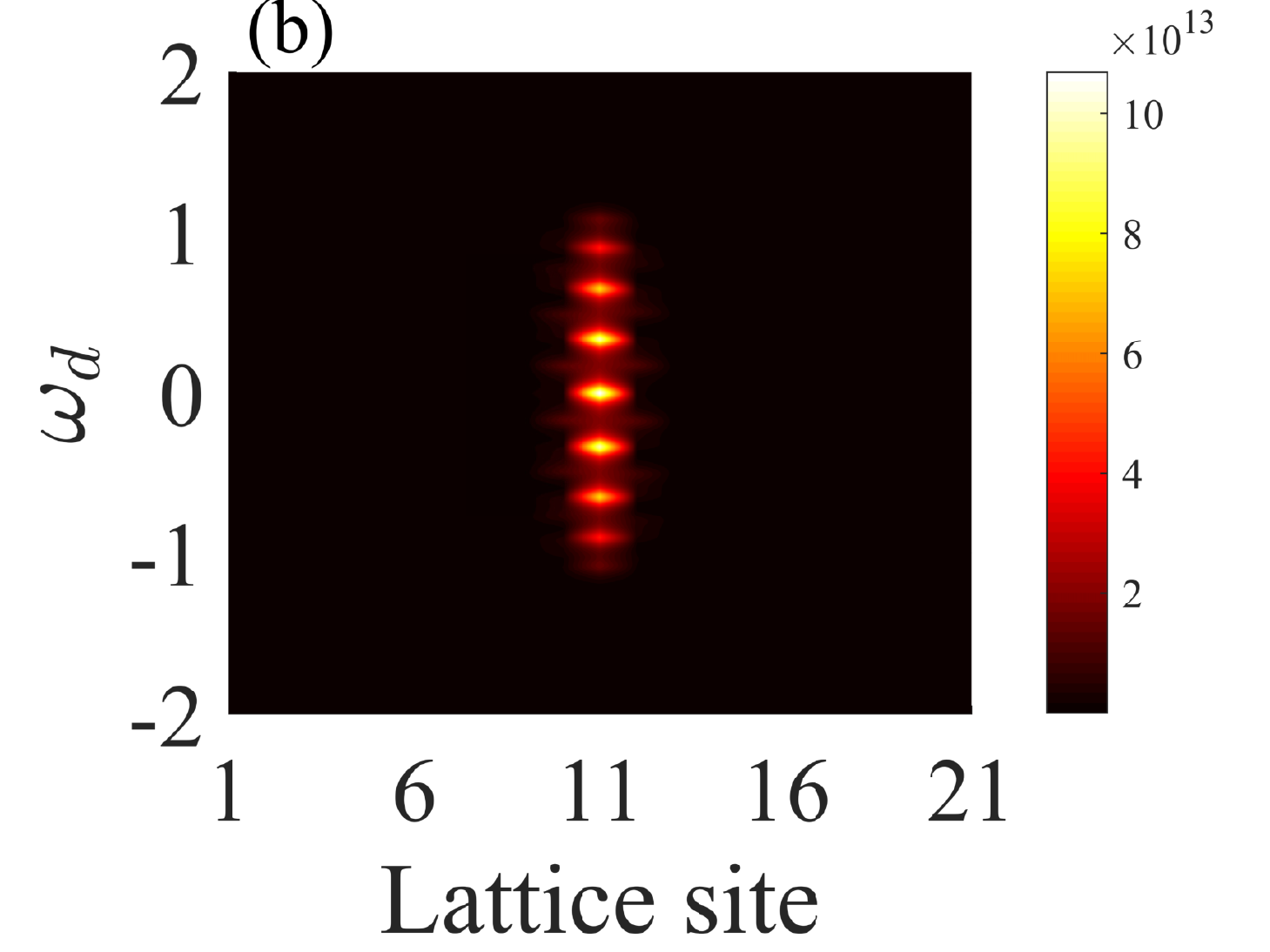}}
	
	\subfigure{\includegraphics[width=0.49\linewidth]{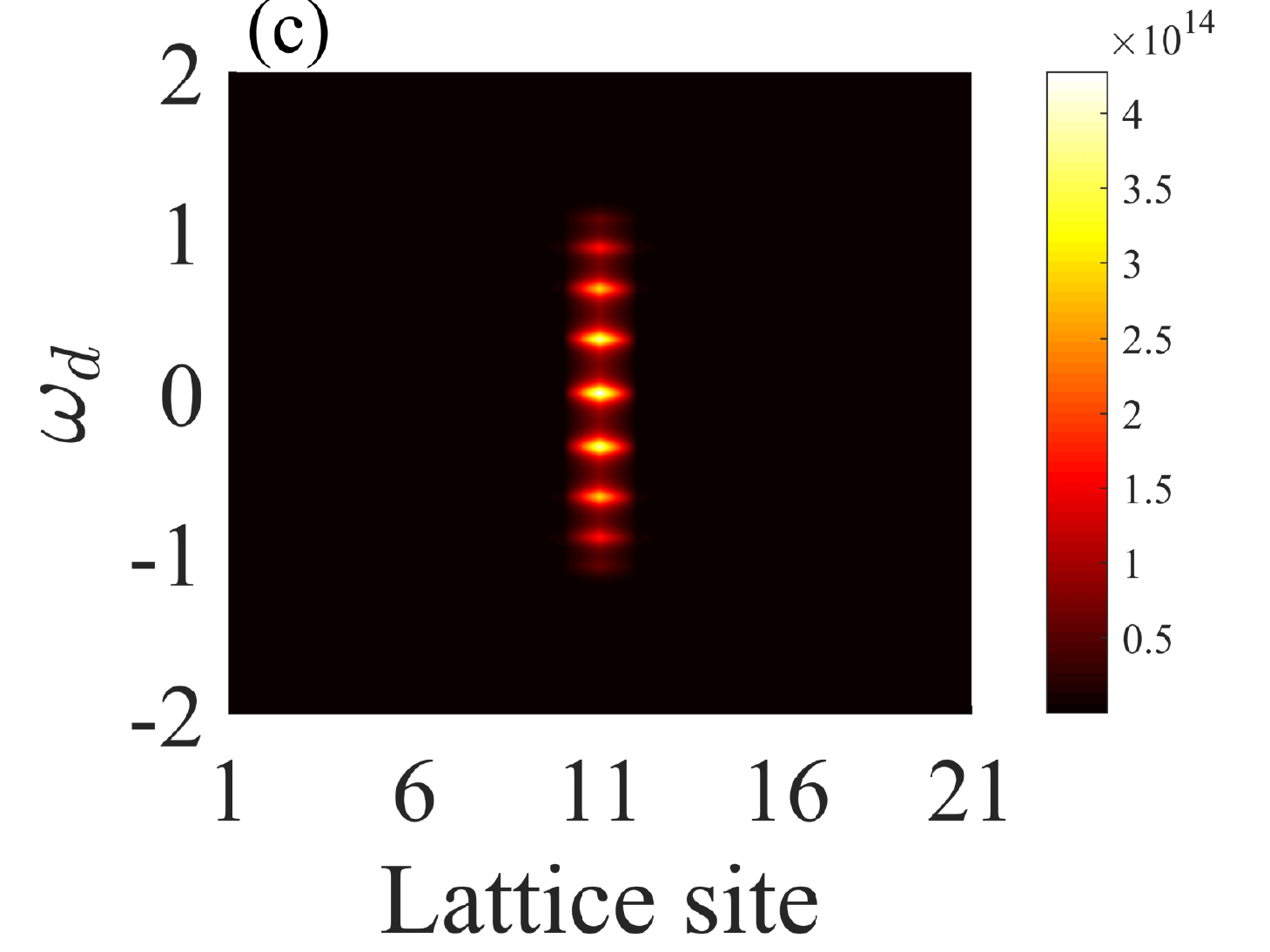}}
	\subfigure{\includegraphics[width=0.49\linewidth]{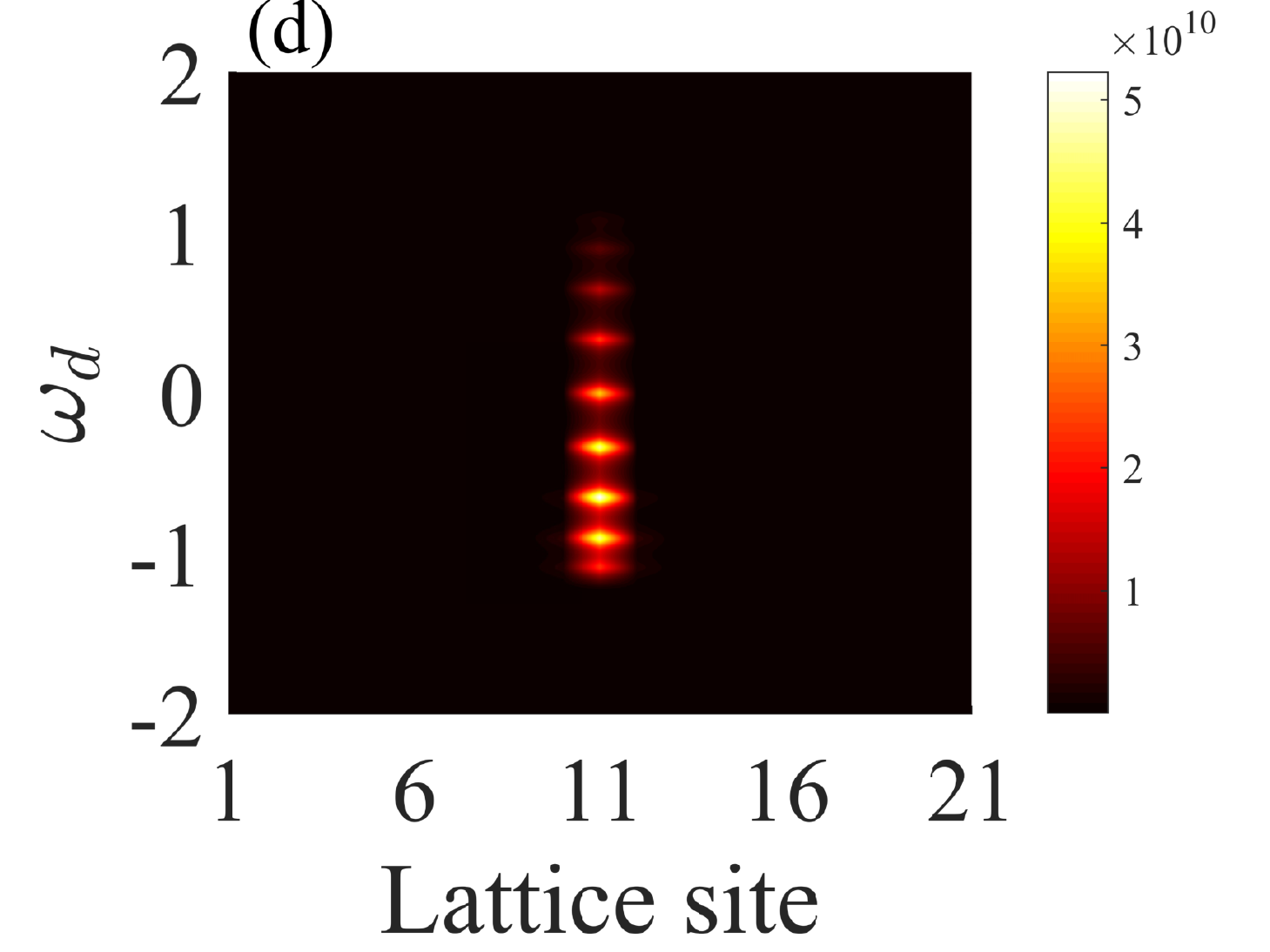}}		
	\caption{The output detection spectra when the external excitations are added on the different resonators. (a) The output detection spectrum when the external excitation is added on the resonator at the interface. (b) The output detection spectrum when the external excitation is added on the first resonator. (c) The output detection spectrum when the external excitations are added on the first and the last resonators at the same time. (d) The output detection spectrum when the external excitations are added on all the resonators. Other parameters take $t_{1}=1$, $\delta=0.8$, and $t_{2}=1$. We set $t_{1}$ as the unit of energy.}\label{fig10}
\end{figure} 

The interface-state laser can further be comprehended via the evolution of the photons initially prepared into the middle resonator $Q$, as shown in Fig.~\ref{fig8}(a). Obviously, the evolution of the photons initially prepared into the middle resonator $Q$ exhibits a bamboo-shape, which is induced by the alternative evolution of the two types of the eigenstates. This kind of distribution of the alternative evolutions with time is equivalent to a pulsed interface-state laser, in which the photons gather into the middle resonator $Q$ intermittently. To implement the interface-state laser easier in experiment, we expect that the interface-state laser can also be achieved when the photons are initially prepared into other resonators. Thus, we prepare the initial state when the first resonator is excited, and we plot the evolution of the initial state, as shown in Fig.~\ref{fig8}(b). Indeed, the pulsed interface-state laser can be achieved with the developing of time. Especially, the pulsed interface-state laser can be reappeared perfectly when the two ends resonators are excited simultaneously, as shown in Fig.~\ref{fig8}(c). Besides, we also investigate the interface-state laser when all resonators are excited initially, as shown in Fig.~\ref{fig8}(d). The numerical results show that the photons are better gathered into the middle resonator $Q$, which is different from the previous case of the pulsed interface-state laser.

Now, we devote to reveal the effects of the on-site defects on the interface-state laser in the present non-Hermitian resonator array, since it is important to the realistic applications in experiment. We find that, when the middle resonator $Q$ is excited, the defect added on the middle resonator $Q$ makes the original pulsed interface-state laser to be a non-pulsed interface-state laser, as shown in Fig.~\ref{fig9}(a). The reason of the above phenomenon is that the large enough on-site defect makes the middle resonator decouple from the resonator array directly due to the absence of the protection of the energy gap. Also, we investigate the case that the on-site defects are separately added on the second and the third resonator when the first resonator is excited initially, as shown in Figs.~\ref{fig9}(b) and~\ref{fig9}(c). We find that the non-pulsed interface-state laser and the pulsed interface-state laser can still be achieved without the influence of the on-site defects. We stress that this immunity to the defects is induced by the nonreciprocal couplings of the non-Hermitian resonator array~\cite{longhi2015non,longhi2015robust,gangaraj2018topological} instead of the protection of the topology of the system. Dramatically, when the defect is added on the lateral resonator of the middle resonator $Q$, we find that the photons gather into the two adjacent resonators alternatively with the time developing, as shown in Fig.~\ref{fig9}(d). It means that we can achieve two different kinds of complementary pulsed interface-state laser in different time regions. 

Before conclusion, we give an analysis of the experimental feasibility. As mentioned above, the interface state laser can be realized when an arbitrary resonator is excited. Thus, in experiment, we can add the external driving on the different resonators, and then the photons will intermittently gather towards the resonator at the interface in a proliferative way due to the non-Hermitian evolution. For example, when we use an external driving with a range of frequency to scan the resonator at the interface, the simulation results show that the photons intermittently accumulate into the resonator at the interface with the varying of the scanning frequency, as shown in Fig.~\ref{fig10}(a). Similarly, when other resonators are excited by the external driving, the photons still mainly gather into the resonator at the interface, as shown in Figs.~\ref{fig10}(b)-\ref{fig10}(d). It means that the pulsed laser at the interface of the resonator array can be realized via the external excitation. Furthermore, we note that a similar interface-state excitation scheme based on the Hermitian micro-resonator chains has been reported in Ref.~\cite{charles2015selective}, in which the defect induced interface state can be excited only when the defect resonator at the interface is excited. Different from the above Hermitian scheme, our scheme ensures that the interface-state laser can be realized when an arbitrary resonator is excited, which greatly decrease the limitation of the experimental realization. At the same time, there is also a non-Hermitian topological laser scheme based on the cavity array plane with the gain and loss being proposed in Refs.~\cite{bandres2018topological,harari2018topological}, in which the photons can propagate along the boundary of the two dimensional cavity array plane and realize the proliferation induced by the gain material. Compared with the above non-Hermitian cavity array plane scheme, our scheme of realizing the interface-state laser has two prominent advantages. One advantage is that our scheme is based on a 1D resonator chain, which can keep its simplest structure in space. Another advantage is that the proliferation of the photons in our scheme is dependent on its intrinsic non-Hermiticity rather than the gain material. Further, our scheme can also induce the topologically trivial and nontrivial interface-state lasers at the same time via designing the appropriate nonreciprocal coupling configurations. Thus, our scheme provides a feasible path to induce interface-state laser and provides significant convenience in experiment.

\section{\label{sec.4}Conclusions}
In conclusion, we have proposed a scheme to achieve the interface-state laser based on a non-Hermitian micro-resonator array, which is composed by two coupled resonator chain via an auxiliary resonator. We find that, when the couplings between the two resonator chains and the auxiliary resonator is weak enough, the photons mainly gather into some certain resonators near the interface if the different resonators are excited initially. The gathering of the photons, towards the certain resonators, has many potential applications in the photon storage device and the laser generator device. Specially, we investigate the case that the resonator array takes the non-Hermitian topological trivial configuration, in which the pulsed interface-state laser can be achieved corresponding to the excitation of the photons in an arbitrary resonator. Also, we reveal that the large enough on-site defect has no effect on the interface-state laser in some cases. The reason is that the nonreciprocal couplings of the resonator array induce the robust photons transmission, which is different from the usual topological protection. Our scheme provides a feasible and appealing method to investigate the interface-state laser based on the macro-resonator array both in theory and experiment.

\begin{acknowledgments}
This work was supported by the National Natural Science Foundation of China under Grant Nos.
61822114, 11874132, 61575055, and 11575048, and the Project of Jilin Science and Technology Development for Leading Talent of Science and Technology Innovation in Middle and Young and Team Project under Grant No. 20160519022JH.
\end{acknowledgments}




\end{document}